%% file: ms.tex
\documentclass[12pt,preprint]{aastex}

\newcommand{\vb}{$V$}
\newcommand{\ib}{$i'$}
\newcommand{\zb}{$z'$}
\newcommand{\alp}{$\alpha$ }

\newcommand{\sol}{$_{\odot}$ }
\newcommand{\super}[1]{$^{#1}$}
\newcommand{\sub}[1]{$_{#1}$}
\newcommand{\lya}{Ly$\alpha$}
\newcommand{\ha}{H$\alpha$}
\def\farcs{\hbox{$.^{\prime\prime}$}}
\def\arcs{\hbox{$^{\prime\prime}$}}

\shorttitle{\lya Galaxies at z = 4.4}
\shortauthors{Finkelstein et al.}

\begin{document}
\title{Effects of Dust Geometry in Lyman Alpha Galaxies at z = 4.4\altaffilmark{1}}

\author{Steven   L.    Finkelstein\altaffilmark{2},   James   E.    Rhoads,   Sangeeta
Malhotra, Norman Grogin\altaffilmark{3} \& Junxian Wang\altaffilmark{4}}   
\affil{$^{1}$  Based in part on observations taken at the Cerro Tololo Inter-American Observatory.}
\affil{$^{2}$  Department  of  Physics,  Arizona  State
University,  Tempe, AZ  85287;  stevenf@asu.edu} 
\affil{$^{3}$  School of Earth and Space Exploration,  Arizona  State
University,  Tempe, AZ  85287} 
\affil{$^{4}$ Center for Astrophysics, University of  Science and Technology  of China, Hefei,  Anhui 230026, China}

\begin{abstract}
  Equivalent  widths  (EWs)  observed  in  high-redshift  Lyman  alpha
  galaxies  could be  stronger than  the EW  intrinsic to  the stellar
  population   if  dust  is   present  residing   in  clumps   in  the
  inter-stellar  medium (ISM).   In this  scenario,  continuum photons
  could  be extinguished  while the  \lya~photons would  be resonantly
  scattered  by  the  clumps,  eventually  escaping  the  galaxy.   We
  investigate this  radiative transfer scenario  with a new  sample of
  six \lya~galaxy candidates  in the GOODS CDF-S, selected  at z = 4.4
  with  ground-based  narrow-band  imaging  obtained at  CTIO.   Grism
  spectra from the  {\it HST} PEARS survey confirm  that three objects
  are at z = 4.4, and  that another object contains an active galactic
  nuclei (AGN).  If we assume  the other five (non-AGN) objects are at
  z =  4.4, they have rest-frame EWs  from 47 -- 190  \AA.  We present
  results of stellar population studies of these objects, constraining
  their rest-frame UV with {\it HST} and their rest-frame optical with
  {\it Spitzer}.   Out of  the four objects  which we  analyzed, three
  objects were best-fit to contain stellar populations with ages on the
  order of 1 Myr and stellar  masses from 3 -- 10 $\times$ 10\super{8}
  M\sol, with dust in the amount  of A\sub{1200} = 0.9 -- 1.8 residing
  in a  quasi-homogeneous distribution.   However, one object  (with a
  rest EW  $\sim$ 150  \AA) was best  fit by  an 800 Myr,  6.6 $\times$
  10\super{9} M\sol  stellar population with a smaller  amount of dust
  (A\sub{1200} = 0.4) attenuating  the continuum only.  In this object,
  the EW  was enhanced  $\sim$ 50\% due  to this dust.   This suggests
  that large EW \lya~galaxies  are a diverse population.  Preferential
  extinction  of  the  continuum  in  a clumpy  ISM  deserves  further
  investigation as  a possible cause of the  overabundance of large-EW
  objects that have been seen in narrow-band surveys in recent years.
\end{abstract}


\section{Introduction}
Over the  last decade, numerous surveys have  been conducted searching
for galaxies with strong  Lyman alpha emission (\lya~galaxies) at high
redshift  (e.g., Rhoads  et al  2000, 2004;  Rhoads \&  Malhotra 2001;
Malhotra \& Rhoads 2002; Cowie \&  Hu 1998; Hu et al 1998, 2002, 2004;
Kudritzki et al 2000; Fynbo, Moller, \& Thomsen 2001; Pentericci et al
2000; Ouchi et al 2001, 2003, 2004; Fujita et al 2003; Shimasaku et al
2003, 2006;  Kodaira et  al 2003;  Ajiki et al  2004; Taniguchi  et al
2005;  Venemans  et  al  2002,  2004).  Many  of  these  studies  have
discovered \lya~galaxies with  large rest-frame \lya~equivalent widths
(EWs;  Kudritzki et  al.  2000;  Malhotra \&  Rhoads  2002; Dawson  et
al. 2004, 2007; Shimasaku et  al. 2006).  Since it was first suggested
40 years ago (Partridge and  Peebles 1967) that \lya~emission would be
an indicator of star formation in the first galaxies during formation,
it has been  thought that strong \lya~emission would  be indicative of
copious star formation.   However, the ratio of high-low  EWs found in
many  of these  surveys is  too high  to be  explained by  a so-called
“normal”  stellar  population  with  a Salpeter  (1955)  initial  mass
function (IMF) and a constant  star formation rate (SFR).  This normal
stellar population has  a maximum EW of 260  \AA~at 10\super{6} years,
settling towards 95 \AA~by 10\super{8}years.  Thus something is causing
the  EWs in  many  of the  observed  \lya~galaxies to  be higher  than
normal.

There are three possible causes  for a stronger than expected \lya~EW,
or a larger high-low EW ratio.  Strong \lya~emission could be produced
via  star  formation if  the  stellar  photospheres  were hotter  than
normal.  This  could happen in extremely low  metallicity galaxies, or
galaxies  which have their  stellar mass  distributed via  a top-heavy
IMF, forming  more high-mass stars than normal.   More high-mass stars
would result  in more ionizing  photons, which would thus  create more
\lya~photons  when they  interact with  the local  interstellar medium
(ISM). Low-metallicity  or top--heavy  IMFs are possible  in primitive
galaxies, which  are thought  to contain young  stars and  little dust
(Ellis  et al.   2001; Venemans  et al.  2005; Gawiser  et  al.  2006,
Pirzkal et al.  2007; Finkelstein et al. 2007).

Active  galactic  nuclei  (AGNs)   can  also  produce  large  \lya~EWs.
However, the  lines in Type I AGN  are much broader than  the width of
the narrow-band  filter, and none of  the accompanying high-ionization
state emission lines have  been detected in many \lya~galaxies (Dawson
et al.   2004).  While the lines in  Type II AGN are  narrower, a deep
{\it Chandra} exposure in the  LALA fields showed no significant X-ray
flux,  even when  individual galaxies  were stacked  (Malhotra  et al.
2003;  Wang et al.   2004).  Wang  et al.   determined that  less than
4.8\% of  their sample could be  possible AGNs.  Thus  while one could
never  entirely  rule   out  AGNs  from  their  sample,   if  one  uses
well-thought out selection criteria, one could be confident that there
are  at  most  a  small   number  of  AGN  interlopers.   In  general,
narrow-band selection techniques usually result in a low AGN fraction.

While \lya~galaxies are historically  thought to be primitive and dust
free, there  is a scenario involving  dust which could  cause an older
stellar  population to  exhibit  a  strong \lya~EW.   If  the dust  is
primarily  in cold,  neutral clouds  with a  hot,  ionized inter-cloud
medium  the  \lya~EW  would   be  enhanced  because  \lya~photons  are
resonantly scattered (Neufeld 1991;  Hansen \& Oh 2006).  Assuming the
clouds  are  thoroughly mixed  with  neutral  hydrogen, \lya~would  be
preferentially  absorbed  by the  hydrogen  at  the  surface of  these
clouds, while  continuum photons would be more  likely penetrate these
clouds more deeply before they get absorbed by dust.  The \lya~photons
would  likely   be  re-emitted  quickly,   proceeding  to  effectively
``bounce'' around  the ISM by  being absorbed and re-emitted  right on
the  surface of  these clouds.   Thus \lya~photons  would have  a much
greater chance  of escaping the  galaxy than continuum  photons, which
would suffer a much greater  chance of extinction.  In this scenario a
strong \lya~EW  could be  observed in a  galaxy with an  older stellar
population and a  clumpy dusty interstellar medium.  Our  goal in this
paper is to research the likelihood that this scenario is occurring in
a  sample of  \lya~galaxies. 

In this paper we  assume Benchmark Model cosmology, where $\Omega_{m}$
= 0.3,  $\Omega_{\Lambda}$ =  0.7 and H$_{0}$  = 0.7 (c.f.  Spergel et
al. 2007).  All  magnitudes in this paper are  listed in AB magnitudes
(Oke \&  Gunn 1983).  In Section  2 we discuss our  data reduction and
sample.  In  Section 3 we  discuss our stellar population  models.  In
Section 4 we detail our results,  and they are discussed in Section 5.
We present our conclusions in Section 6.

\section{Data Handling}

\subsection{Observations}
In order  to study \lya~galaxies at z  = 4.4, we obtained  a 2.75 hour
exposure in  the 6563 \AA~(\ha) narrow-band filter  (NB656) in October
2006 using  the MOSAIC  II CCD  imager on the  Blanco 4m  telescope at
Cerro Tololo  Inter-American Observatory  (CTIO).  MOSAIC II  has eight
chips over a 36$^{\prime}$  $\times$ 36$^{\prime}$ field of view, with
each  pixel covering  0\farcs27 on  the  sky.  In  our previous  study
(Finkelstein et  al.  2007), we obtained  ground-based broad-band data
of   regions  with  pre-existing   narrow-band  data.    However,  the
broad-band  data was  not nearly  deep enough  to constrain  the dusty
scenario.

We  observed  the  Great  Observatories Origins  Deep  Survey  (GOODS;
Giavalisco  et al.   2004) Chandra  Deep  Field --  South (CDF--S;  RA
03:31:54.02,  Dec  -27:48:31.5  (J2000))  to  take  advantage  of  the
extremely deep {\it HST}/Advanced Camera for Surveys (ACS; Ford et al.
1998)  and {\it  Spitzer}/Infrared Array  Camera (IRAC;  Fazio  et al.
2004) broad-band data publicly available.  In addition, the photometry
and  astrometry of  these  broad-band data  have  been extremely  well
determined, further reducing errors  in our results.  The data consist
of the following broad-band filters: F435W (B), F606W (V), F775W (i'),
F850LP  (z'), 3.6  $\mu$m (Channel  1),  4.5 $\mu$m  (Channel 2),  5.8
$\mu$m (Channel 3) and 8.0 $\mu$m (Channel 4).

\subsection{Data Reduction}

The CTIO  data were reduced with  IRAF\footnote[1]{IRAF is distributed
  by  the National  Optical Astronomy  Observatory (NOAO),  which is
  operated  by  the  Association   of  Universities  for  Research  in
  Astronomy,  Inc.\  (AURA)   under  cooperative  agreement  with  the
  National  Science  Foundation.}   \citep{tody86,tody93},  using  the
MSCRED  \citep{valtod,val98} reduction  package, following  the method
set  forth in  Rhoads et  al.  (2000, 2004).   First, we  performed the
standard   image  reduction  steps   of  overscan   subtraction,  bias
subtraction and  flat--fielding.  Cross-talk was  also removed between
chip pairs  sharing readout electronics.   We derived a  supersky flat
from  the  science  data,  and   used  this  to  remove  the  residual
large--scale imperfections  in the sky.  The  world coordinate systems
(WCS) of  individual frames were  adjusted by comparing the  frames to
the astrometry from the USNO--A2.0 catalog.  Cosmic rays were rejected
using  the  algorithm  of  Rhoads  (2000) and  satellite  trails  were
manually flagged  and excluded from the final  stacked image.  Weights
for  image  stacking were  determined  using  the ATTWEIGHT  algorithm
(Fischer \& Kochanski  1994), with the task {\it  mscstack} being used
to  make the  final stack.   See  Rhoads et  al.  (2000)  and Wang  et
al. (2005) for further details on data reduction.

\subsection{Object Extraction}
In order to select \lya~galaxies in our field, we used the GOODS CDF-S
ACS v1.0 catalogs, obtained from the GOODS website.  Because the GOODS
ACS  data  are much  deeper  and have  a  higher  resolution than  our
narrow-band  data, we chose  to extract  sources from  our narrow-band
data using  point-spread function  (PSF) fitting routines  rather than
aperture photometry.  We  were able to do this  because typical high-z
\lya~galaxies are not resolved from  the ground (Pirzkal et al. 2007).
We  performed PSF  fitting  using the  routines  in the  IRAF port  of
DAOPhot (Stetson 1987).

We  chose the  positions to  fit  the PSF  based on  the positions  of
sources in the GOODS catalogs.  The objects we were looking for were z
$\approx$ 4.4 \lya~galaxies, so they  should be B-band dropouts with a
narrow-band  excess.   We  used  the  B-dropout  color  criteria  from
Giavalisco et al.  (2004) with  the magnitudes from the GOODS catalogs
in order to find out which objects were Lyman Break Galaxies (LBGs) at
z $\sim$ 4.4.  The number  counts of the \vb~magnitudes of all objects
in the  catalog turned over just  brighter than 27th  magnitude, so we
chose to  only fit the PSF to  objects with \vb~$\leq$ 27  in order to
avoid  selecting objects beyond  the completeness  limit of  the GOODS
dataset.

The input  list to the  PSF-fitting routine consisted  of the x  and y
positions in  the narrow-band  image corresponding to  the RA  and DEC
coordinates of  the objects we wanted  to fit from  the GOODS catalog.
To  be sure  the  coordinates  matched up  correctly,  we performed  a
transformation to  correct for any  WCS distortion between the  WCS of
the narrow-band image and  the GOODS catalog.  The PSF-fitting routine
was then given  this corrected list of x and  y coordinates along with
the PSF image computed from the narrow-band image.  The outputs were a
file of the object magnitudes along with a subtraction image (an image
showing the residuals left when  the fitted PSFs were subtracted out).
On  inspecting  this  image,  we  found that  with  the  exception  of
extremely bright stars or extended objects, this method of PSF fitting
extracted the objects very well.

In  order to  calculate  the  zero-point of  the  H$\alpha$ image,  we
plotted a  color-magnitude diagram of  H$\alpha$ - \vb~vs.   \vb.  The
H$\alpha$ filter is contained  entirely within the \vb~filter, so that
the  mean  colors  of  these  objects  should be  zero.   We  fit  the
zero-point  using only  unsaturated stars,  and find  a  zero-point of
27.55  mag for 1  count in  the final  image.  The  DAOPhot photometry
errors  imply  a  5$\sigma$   limiting  magnitude  of  24.90  for  the
narrow-band image.

\subsection{Lyman \alp Galaxy Selection}
We  find z  $\approx$ 4.4  galaxy  candidates by  seeing which  B-band
dropouts   exhibit  a   narrow-band  excess.    Figure  1   shows  the
color-magnitude  diagram   for  all  of  the   objects  we  extracted,
highlighting  the  B-band  dropouts,  along with  the  initial  galaxy
candidates.   There were  11 initial  candidates, and  six of  them we
eventually determined  to be  true candidates after  visual inspection
showed that they were  real, uncontaminated objects.  The five objects
which were rejected at this stage were either contaminated by a nearby
star or were not visible above the noise in the NB656 image.

We  performed a  number of  tests  to check  the validity  of our  six
candidates.  First, we looked for other B-dropouts with similar V-band
magnitudes  to each of  the six  candidates, and  then checked  to see
whether   these  somewhat   randomly  selected   objects   showed  any
narrow-band excess,  which none  of them did.   We then looked  at the
subtraction  image.  With  a  perfect PSF  extraction,  we should  see
nothing at  the location of  our candidates.  Indeed,  when inspecting
these subtracted  areas, nothing stands out from  the noise.  Finally,
we took a  mosaicked image of the GOODS V-band  tiles and convolved it
to  the seeing of  the ground-based  \ha~image ($\sim$  0\farcs9).  We
then subtracted this  image from the \ha~image, and  checked to insure
that the six candidates showed a narrow-band excess, which five out of
the  six did.   The object  that did  not show  an excess  upon visual
inspection is near a bright star (object 5), and also has the smallest
computed excess,  which could explain  why it doesn't stand  out above
the noise in the  \ha~- V image. Figure 2 shows stamps  of each of the
six candidates, along  with the results of some  of the methods listed
above.  Figure 3 shows stamps of each candidate in each of the \ha, B,
V, i', z', 3.6$\mu$m and 4.5$\mu$m  bands.  Table 1 details the EWs of
each object, while Table 2 list the magnitudes of each object for each
filter we used.

\subsection{Spectroscopic Confirmations}

While  we  have  assumed  that  our narrow-band  filter  has  selected
\lya~galaxies  at  z $\approx$  4.4,  it  could  also pick  up  [OIII]
emitters  at  z  $\sim$ 0.3  and  [OII]  emitters  at z  $\sim$  0.75.
Ideally,  we would  like to  be able  to see  a \lya~line  in  a given
spectrum to make a confirmation,  but lacking that we can still confirm
a redshift based  off of a Lyman break.   The PEARS (Probing Evolution
and  Reionization Spectroscopically; PI  Malhotra) {\it  HST} Treasury
Program  (\#10530) has  obtained ACS  grism spectroscopy  of  $\sim$ a
third of the  solid angle of the CDF--S via 5  ACS pointings.  Four of
our candidates were observed in PEARS.

Based off  of the positions of  observed emission lines  and breaks in
the observed  spectra, we have been able  to spectroscopically confirm
objects 1, 2 and 6 to be at z $\approx$ 4.4.  While Object 4 is not in
any  of the PEARS  fields, it  is covered  by the  GOODS--MUSIC Survey
(Grazian et  al.  2006), and it  was calculated to  have a photometric
redshift of 4.42.  For completeness, we list the photometric redshifts
for Objects 1, 2 and 6, which are 4.24, 4.44 and 4.36 respectively.

In order to search for AGNs  in our sample, we checked the Chandra Deep
Field --  South dataset (Giacconi et  al.  2001), which  consists of a
deep  {\it Chandra X-ray  Observatory} exposure  in the  GOODS CDF--S.
One object was  detected in the Chandra data (Object  3), and we found
in the literature that it is a known [C IV] $\lambda$1549 emitter at z
= 3.19  (Szokoly et al.  2004; Xu et  al. 2007).  This object  was not
included in our analysis.

\subsection{Broad-band Photometry}
For  our analysis,  we used  the updated  version of  the  GOODS CDF-S
catalog  (v1.9; Giavalisco  et al.   2007  in prep)  which has  deeper
observations in  the \ib~and  \zb, resulting in  lower error  bars for
those two bands.  As an external check on the photometry errors in the
GOODS catalog, we used the  Hubble Ultra Deep Field (HUDF; Beckwith et
al.  2006) catalog.  The HUDF is much deeper than the v1.9 data in the
same regions, thus any errors would be dominated by the v1.9 data.  We
matched  the  HUDF  and  v1.9  catalogs, then  plotted  the  magnitude
difference between  the two  vs. the  v1.9 data for  each of  the four
bands  (B, V,  \ib~\&  \zb).   We characterized  this  error as  being
1$\sigma$  of the  spread in  the  magnitude difference  in a  certain
magnitude slice.  This error was small  ( $<$ 0.1 mag in our magnitude
regime), but in the cases where it was larger than the v1.9 photometry
error\footnote[2]{The 1$\sigma$  spread in the  v1.9 photometry errors
  was  0.020,  0.027  and  0.023  mag  for the  V,  i'  and  z'  bands
  respectively at 26th magnitude.}, we used this error in its place.

When comparing data from different  telescopes to models, we needed to
be sure  that the magnitudes we  were using from  our narrow-band, ACS
and IRAC data  meant the same thing.  In  the narrow-band, we obtained
fluxes from PSF fitting, so then the magnitudes from these fluxes were
total  magnitudes.  For the  GOODS/ACS data,  we used  the MAG$\_$AUTO
magnitudes, which  are known to  be 5\% fainter than  the total
magnitude (Bertin 2006).  Thus we corrected the GOODS/ACS data, adding
on a factor of 5\% to the flux to make them total magnitudes.

The GOODS {\it Spitzer}/IRAC data do  not yet have a public catalog, but a
catalog has been  created using the TFIT software  package (Laidler et
al.  2007).  TFIT  uses {\it a priori} knowledge  of spatial positions
and  morphologies from  a higher-resolution  image (in  this  case the
GOODS ACS \zb~image) to construct object templates and fit them to the
lower  resolution  image.   However,   this  used  the  ACS  catalog's
MAG$\_$ISO parameter as  a guideline for the radius  of the galaxy, so
in order to find the total  magnitude, we added to the IRAC magnitudes
the difference  between MAG$\_$ISO and MAG$\_$AUTO for  our objects in
the ACS  \zb~catalog.  We  then added an  additional factor of  5\% to
correct the IRAC magnitudes to be total magnitudes.  We now had total
magnitudes for our objects in nine bands, from $B$ -- 8.0 $\mu$m.

\section{Modeling}

\subsection{Stellar Population Models}
We have derived the physical properties of our candidate \lya~galaxies
by  comparing them to  stellar population  models, using  the modeling
software of  Bruzual \& Charlot (2003; hereafter  BC03).  Comparing to
models allows  one to choose a  variety of physical  parameters to see
which  matches the  observations the  best.  We  attempted to  fit the
following parameters for each galaxy: age, metallicity, star formation
rate (SFR), mass and dust content.   We used a grid of 24 ages ranging
from 1 Myr to  1.35 Gyr (the age of the Universe at  z = 4.4).  We fit
the objects to the full range of metallicity available with BC03, from
0.005 Z\sol -- 2.5 Z\sol.  We modeled exponentially decaying SFRs with
characteristic  decay times ($\tau$)  of 10\super{6},  10\super{7} and
10\super{8}  yr.  We also  approximated an  instantaneous burst/Simple
Stellar Population  (SSP) by using an exponentially  decaying SFR with
$\tau$ = 10\super{3}  yr, and a constant SFR with  $\tau$ = 4 $\times$
10\super{9} yr.  We included dust via the Calzetti dust extinction law
(Calzetti et  al.  1994), which  is applicable to  starburst galaxies,
using  the range  0 $\leq$  A\sub{1200} $\leq$  2.  Galactic  dust was
included from  the method of Schlegel  et al.  (1998) with  a value of
E(B-V) = 0.009 mag.   We include intergalactic medium (IGM) absorption
via the prescription of Madau (1995).

The BC03  models do not tabulate  any emission lines, but  in order to
model the dusty scenario we  needed to include the \lya~emission line.
We  derived the  \lya~line flux  from the  number of  ionizing photons
output for each model,  assuming Case B recombination (see Finkelstein
et  al. 2007  for details).   Some of  our objects  showed  a stronger
3.6$\mu$m flux than  would have otherwise been expected.   At z = 4.4,
the \ha~emission line  would fall in this filter,  so we also included
an \ha~emission line in the models (assuming Case B recombination, the
\ha~line strength is 0.112  $\times$ the \lya~line strength; Kennicutt
1983).

In order to  model the clumpy dust scenario, we  needed to ensure that
the Ly$\alpha$ line did not  suffer dust attenuation.  To do this, the
continuum flux (and \ha~flux) was  multiplied by the Calzetti dust law
before we added in the Ly$\alpha$  flux to the spectrum at the correct
wavelength bin.  In this  case, the continuum suffers dust attenuation
while the \lya~line  does not.  \lya~sits right at  a step function in
the  Madau IGM  treatment, so  the  amount of  attenuation applied  to
\lya~depends  strongly  on  the   exact  wavelength  position  of  the
\lya~line.    The  accepted  interpretation   is  that   the  internal
kinematics  of a given  galaxy will  result in  half of  the \lya~flux
coming out  slightly blue  of the rest  wavelength, and  half slightly
red.   This results in  the characteristic  asymmetric profile  of the
\lya~observed in many  spectra, where the blue side  is truncated.  We
approximate  this interpretation  by attenuating  the \lya~line  to be
$\sim$ one half of its original (post-dust where applicable) flux.

In order to go from the  BC03 output flux to bandpass averaged fluxes,
we used the method outlined by  Papovich et al. (2001).  In short, we
took    the    output   from    BC03,    and    converted   it    from
L\sub{\ensuremath{\lambda}}  into  L\sub{\ensuremath{\nu}},  and  then
from  L\sub{\ensuremath{\nu}} into  F\sub{\ensuremath{\nu}}  (units of
erg       s\super{-1}      cm\super{-2}       Hz\super{-1})      using
F\sub{\ensuremath{\nu}}   =   (1   +  z)   L\sub{\ensuremath{\nu}}   /
(4$\pi$d{\super2}\sub{L})  to  do  the  conversion  and  redshift  the
spectrum.  This flux was  then multiplied by the transmission function
for  a  given bandpass  (including  the  filter  transmission and  the
quantum  efficiency   of  the  detector),  and   integrated  over  all
frequency.          The         bandpass         averaged         flux
$\left<f_{\ensuremath{\nu}}\right>$ is  this result normalized  to the
integral of the transmission  function.  AB magnitudes \citep{oke} for
the models were then computed.

\subsection{Dust Effects on the \lya~EW}

The scenario  we are  probing observationally is  whether or  not dust
could enhance  the \lya~EW.  Traditionally,  it has been  thought that
\lya~galaxies could possibly be too primitive to have formed dust yet.
However, some evidence  does exist that would enable  a galaxy to form
dust  quickly.  Massive  stars evolve  on a  very short  timescale, so
after a few Myr, a galaxy could begin to have some heavier elements in
its  ISM.  Also, supersolar  metallicities and  CO emission  have been
seen in quasars at z $\sim$ 6 (Pentericci et al. 2002; Bertoldi et al.
2003)  and  \lya~emission has  been  seen  in sub-millimeter  selected
galaxies (Chapman et al. 2004).

However, even if \lya~galaxies did have dust, it has been thought that
it  would vastly attenuate  the \lya~flux  (Meier \&  Terlevich 1981).
\lya~photons have long  scattering path-lengths, making them extremely
vulnerable  to dust  attenuation, assuming  a uniform  distribution of
dust.   It is possible  that dust  could be  geometrically distributed
such  that the \lya~photons  are attenuated less than  the continuum.
The most likely of these scenarios  involves an ISM  with clumpy dust
clouds  thoroughly mixed  with  neutral hydrogen  with an  inter-cloud
medium which is tenuous and ionized (Neufeld 1991; Hansen \& Oh 2006).
Evidence  for  the possible  existence  of  this  scenario also comes  from
Giavalisco et al.   (1996), who find a lack  of correlation between UV
slope and \lya~EW, meaning that  a UV slope indicative of dust doesn't
necessarily mean a  low \lya~EW.  They conclude that  the ISM in their
sample of galaxies is highly  inhomogeneous, and that the transport of
\lya~photons is primarily  governed by the geometry of  the ISM rather
than the amount of dust.

This is intriguing because if \lya~galaxies had a dusty, inhomogeneous
ISM then it is possible that the continuum photons were attenuated and
the \lya~photons weren't.  Then the  observed EW would be greater than
the  EW intrinsic  to the  underlying stellar  population by  a factor
relating to the amount of dust present.  To better understand this, we
will follow the journey of a  photon through an ISM, starting from the
point when  it is emitted  from a massive  star.  This star  will emit
numerous  ionizing  photons, 2/3  of  which  will become  \lya~photons
assuming Case B recombination.  When a \lya~photon encounters a region
of dust mixed with neutral hydrogen, it has a very high probability of
being absorbed by the first hydrogen atom it encounters.  It will then
be re-emitted  in a  random direction, with  an equal chance  of being
sent  back the  way it  came  as being  sent forward  in its  original
direction  of  travel.   If  the  ISM is  uniformly  distributed,  the
\lya~photon  will only  move a  short distance  before it  is absorbed
again, making  it take  a very  long time for  a given  \lya~photon to
escape the galaxy.  If the ISM is composed of clouds in a nearly empty
inter-cloud medium, then the 50\%  of the time that the \lya~photon is
emitted back out  of the clump, it will travel  a long distance before
it encounters another clump, vastly increasing the probability that it
will escape  the galaxy.   Even if it  is re-emitted in  the direction
further  into  the clump,  it  will still  be scattered  close to  the
surface, so it still has a chance to escape.

However, continuum photons in  a homogeneously distributed ISM will be
absorbed  by dust  as they  travel through  the ISM,  but  since their
wavelengths are continuously distributed  they will essentially not be
affected by the presence of  neutral hydrogen.  The same holds true in
a  clumpy ISM.   A continuum  photon will  be able  to  ``bypass'' the
hydrogen in the clump until it encounters a dust grain, at which point
it could either  get scattered or absorbed.  If  the continuum photons
are  absorbed, it will  eventually be  re-emitted, but  its wavelength
will have changed  to the far-IR.  The net effect of  this is to lower
the amount of continuum flux escaping the galaxy (as well as to redden
the  flux  due  to  those  photons which  are  scattered  rather  then
absorbed).  Thus, a  clumpy ISM could result in  a higher \lya~EW than
the underlying stellar population.

It  has been  generally believed  that \lya~galaxies  are  very young,
primitive  objects.   If we  find  proof  that  an object  shows  dust
enhancement of  the \lya~EW, it could  turn out that  some fraction of
\lya~galaxies  known to  exist could  in fact  be older,  more evolved
stellar populations,  changing our view of  where \lya~galaxies belong
in the galaxian zoo.

\subsection{IGM Inhomogeneities}
Although differences in  the IGM can affect the models  in the V band,
we chose to include this band in our fits.  The V-band is complicated,
as at  z =  4.4, it  contains the entire  \lya~forest region  from the
Lyman limit out  to \lya.  To account for  inhomogeneities in the IGM,
we chose to include a flux variance for the model V-band flux (results
in   a  second   parameter   in  the   denominator   for  the   V-band
$\chi$\super{2} term;  see Section 4.1).  To account  for IGM absorption
in our models, we have  attenuated the cumulative flux which comes out
of BC03 by  a factor of exp(-$\tau$\sub{IGM}), such  that the bandpass
averaged flux in the V-band is given by:
\begin{equation}
\left<f_{V}\right> = \int_{\lambda_{1}}^{\lambda_{2}} f_{BC03,\lambda} e^{-(\tau_{\lambda} + \delta\tau_{\lambda})} d\lambda
\end{equation}
where  f$_{BC03,\lambda}$ is  the initial  output of  the  models, and
$\delta\tau_{\lambda}$ is the 1$\sigma$ error in the IGM optical depth
due to  inhomogeneities among  different lines-of-sight.  We  can then
break this up  into a sum over discrete intervals, and  calculate the
variance, which for a small $\delta\tau_{\lambda}$ is:
\begin{equation}
\sigma_{\left<f_{V}\right>}^{2} = \sum_{n=1}^{N} f_{BC03,\lambda}^{2} \left(e^{-\tau_{\lambda}}\right)^{2} \left(\delta\tau_{\lambda}\right)^{2} \Delta\lambda^{2}
\end{equation}
where N is  the number of bins  we chose to break the  V-band into. We
approximate this as
\begin{equation}
\sigma_{\left<f_{V}\right>}^{2} = N \left(\delta\tau_{\lambda}\right)^{2} \frac{F_{V}^{2}}{N^{2}} \left(e^{-\tau_{\lambda}}\right)^{2}
\end{equation}
where F\sub{V}\super{2}  is the total  flux in the bandpass.   The rms
flux in the V-band due to the IGM inhomogeneities is then:
\begin{equation}
\sigma = F_{BC03,V} e^{-\tau} \delta\tau N^{-\frac{1}{2}}
\end{equation}
Thus the fractional error in the model flux which needs to be added in
is $\delta\tau$ $\times$  N$^{-\frac{1}{2}}$.  To make this physically
correct, the bin  size should be equal to  a typical clustering length
for  \lya~forest  clouds  such  that  the  \lya~forest  optical  depth
variations  in adjacent  bins are  essentially  uncorrelated.  Penton,
Shull \&  Stocke (2000) calculated the  two-point correlation function
for \lya~forest clouds.  They found that the correlation drops to zero
at a separation of $\sim$ 11  Mpc.  For z = 4.4, the comoving distance
between the front  and back of the V-band filter is  1205 Mpc, thus we
chose  to  use  110  bins.   The uncertainty  in  the  IGM  absorption
($\delta\tau$) was  found from  Madau (1995).  Extrapolating  from his
example  of  a  z  $\sim$  3.5  galaxy, we  found  an  uncertainty  in
exp(-$\tau$) of $\sim$ 0.35,  which corresponds to $\delta\tau$ $\sim$
= 1.05.  Plugging these values  into the above equations, we find that
we need  to include an error  of $\sim$ 0.1 magnitudes  for the V-band
model fluxes to account for inhomogeneities in the IGM.

\subsection{Equivalent  Widths}  
Model \lya~equivalent  widths were calculated  using the ratio  of the
model flux in  the \lya~wavelength bin to the  average continuum value
on either side of that bin ($\times$ $\Delta\lambda$).  We compare our
model  EW distribution  to  previous studies  (i.e.   Charlot \&  Fall
(1993);  Malhotra  \& Rhoads  (2002;  MR02)).   We  also quantify  the
multiplicative  effect adding  clumpy dust  to the  models has  on the
\lya~EW.

Figure 4  shows the EWs of  our models plotted  vs. stellar population
age  for Z=.02Z\sol  and  three different  star  formation rates:  SSP
(single burst),  $\tau$ =  10\super{7} (exponentially decaying)  and a
constant star formation rate.  We first discuss the results from
the  left-hand  vertical axis,  which  shows  the  EWs from  dust-free
models.  All  of the models  start at some  value ($\sim$ 400  \AA) at
10\super{6} yr  and then decay from  there.  The EWs from  the SSP and
tau models  fall all the way to  zero at approximately 30  and 100 Myr
respectively.    Models    which   are   constantly    forming   stars
asymptotically reach a  constant value of EW at  $\sim$ 120 \AA~due to
the constant replenishment of massive stars.  The upturn at later ages
for the  SSP and tau  models are due  to either hot  horizontal branch
stars and/or planetary nebula nuclei, which will start to dominate the
UV flux  after $\sim$ 10\super{8}  yr.  While technically  the \lya~EW
will be high, the actual UV continuum flux is tiny compared to what it
was at  earlier ages, and thus  objects in the this  stage will likely
not appear in \lya~galaxy samples.   In order to distinguish EW due to
young stars from EW due to faint UV continuum flux, the solid lines in
these plots change  to dashed lines when the  continuum (V--band) flux
falls to 10\% of its value at 10\super{6} yr.

Charlot \&  Fall (1993) computed  the \lya~EW from  stellar population
models using  an earlier  version of the  Bruzual \&  Charlot software
(1993).    Their  models   assumed  solar   metallicity  and   Case  B
recombination.   They computed  the \lya~EW  for multiple  IMF slopes,
upper mass cut-offs and star  formation rates, with their youngest age
being  5  Myr.   For  a  SSP  with  a  near-Salpeter  IMF  (x  =  1.5;
x$_{Salpeter}$ = 1.35) their maximum EW was calculated to be 210 (240)
\AA~for an upper mass cut-off of 80 (120) M\sol, falling to zero at 40
Myr.  The EWs from our SSP model have a maximum value of 400 \AA~(only
$\sim$ 100 \AA~at t = 5 Myr),  falling to zero by $\sim$ 20 -- 30 Myr.
However, their definition  of SSP differs from ours.   We defined ours
as having a  characteristic decay rate of 10\super{3}  yr, while their
burst model had  a time scale of 10\super{7} yr.   We can thus compare
their results  to our tau model  (which has $\tau$  = 10\super{7} yr).
This model has a maximum EW of 400~\AA~($\sim$ 200 \AA~at 5 Myr), much
closer  to  the results  of  Charlot \&  Fall.   For  a constant  star
formation rate,  Charlot \& Fall find  a maximum \lya~EW  of 150 (250)
\AA~falling to a constant value  of 90 (100) \AA~by 10\super{8} yr for
an upper  mass cut-off of 80  (120) M\sol (our upper  mass cut-off was
100 M\sol).  Our  continuously star forming models start  at a maximum
EW of $\sim$ 400 \AA~($\sim$ 220  \AA~at 5 Myr), falling to 120 \AA~by
10\super{8}  yr.   Taking into  account  the  differences between  IMF
slopes, metallicities  and upper mass  cut-offs, we feel  our constant
SFR models  are consistent with those  of Charlot \& Fall.   In a more
recent  study, MR02 did  a similar  analysis using  Starburst99 models
(Leitherer et  al.  1999).   Their Salpeter IMF  model has  a constant
SFR, upper mass  cut-off of 120 M\sol and Z =  0.05Z\sol.  The EW from
this  model had  a  maximum  value of  300  \AA~at an  age  of 1  Myr,
asymptoting  to a  value of  100  \AA~by 100  Myr, both  of which  are
consistent with our models.

When   clumpy  dust  comes   into  the   models,  things   can  change
dramatically.  As  we discussed above, in  our models the  dust is not
attenuating the \lya~line, which results in the \lya~EW being enhanced
as the continuum is suppressed.  Consequently, given an amount of dust
extinction, the  \lya~EW will be enhanced by  a multiplicative factor.
For dust amounts of A$_{1200}$ = 0.5, 1.0, 1.5 and 2.0 this factor is:
1.64, 2.68, 4.40 and  7.21 respectively.  The right-hand vertical axis
in Figure 4 shows what the  EW from the models are with 2.0 magnitudes
of clumpy  dust extinction.  What  this tells us  is that for  a given
model, you can go  out to a larger age while still  keeping a high EW.
For example,  for a  SSP with Z  = 0.02Z\sol,  the EW drops  below 200
\AA~by 3 Myr,  while with 2.0 magnitudes of  clumpy dust, this doesn't
happen until after 10 Myr, over  a factor of 3X longer, making it much
more likely  that this object  would be observed  with a high-EW  in a
narrow-band selected  survey.  Table 3 shows the  maximum possible age
allowed for a stellar population with  EW $<$ 200 \AA~for a variety of
models.

\section{Results}

\subsection{Fitting to Models}   
Using our bandpass averaged fluxes, we were able to fit our objects to
the  BC03 models.   We have  nine possible  bands to  choose  from for
fitting: CTIO-MOSAIC  \ha, {\it HST}/ACS  F435W (B), F606W  (V), F775W
(i'), F850LP (z'), {\it  Spitzer}/IRAC Channel 1 (3.6 $\mu$m), Channel
2 (4.5 $\mu$m), Channel 3 (5.8 $\mu$m) and Channel 4 (8.0 $\mu$m).  In
order to  separate the model-dependent parameters  from other factors,
we elected to fit flux ratios to  the models, and then fit the mass to
the best-fit  model at the  end.  We computed  a flux ratio to  the i'
band for each  band, with a corresponding flux  ratio error propagated
through.

The  $\chi$\super{2}  was  then  computed  for each  model  --  object
combination using:
\begin{equation}
\chi^{2} = \sum_{i} \frac{\left[f_{obj}^{i} - f_{mod}^{i}(t, \tau, A_{1200}, Z)\right]^{2}}{\sigma^{2}(f_{obj}^{i})}
\end{equation}
where $i$  denotes the bandpass,  $f_{obj}$ is the object  flux ratio,
$f_{mod}$ is the model flux ratio and $\sigma$$(f_{obj})$ is the error
in the flux ratio for a  given bandpass, derived from the error in the
flux for each bandpass and the  error in the i'-band flux.  All fluxes
were   in   units  of   f\sub{\nu}   (ergs  s\super{-1}   cm\super{-2}
Hz\super{-1}).

Out of our five \lya~galaxy candidates, we only fit stellar population
models to four  of them.  While we have no reason  to suspect object 5
is  not a real  \lya~emitter, its  proximity to  a very  bright object
makes  its {\it  Spitzer} fluxes  inaccurate.   With only  the ACS  and
narrow-band data to  go on, there are not  enough available pass-bands
to fit this object.  Thus the four  objects which we fit were: 1, 2, 4
and 6.  While we did have data in all nine bands listed above, not all
were used  in our fitting.   Both the IRAC  5.8 $\mu$m and  8.0 $\mu$m
were undetected  in the four objects  we fit, and thus  were not used.
Because the entire  B-band is blue-ward of the Lyman  break, we can be
highly  confident   that  on  any  line-of-sight  there   will  be  no
significant flux  transmitted through the  B-band.  If redshift  was a
free parameter in our models, then we would definitely use the B-band,
as  the  absence  of  flux  would be  a  helpful  redshift  indicator.
However, given  that three of our four  objects were spectroscopically
confirmed (and the fourth has a photo-z)  to be at z = 4.4, we did not
feel it would be worthwhile to introduce redshift as a free parameter.
Thus we have left the B-band out of the fitting.

As a  final step, we  introduced a parameter,  q, which we  called the
geometry  parameter  because  its   value  simulates  the  effects  of
different dust  geometries.  Previously, when  we applied dust  to the
models, we added  in the \lya~flux after the dust  was added, that way
the \lya~flux was not attenuated.   This had a very ``cartoon'' effect
on the models,  meaning that \lya~was either attenuated  by all of the
dust or no dust.  We still do things in the same order as before, only
now  we  multiplied the  \lya~flux  by  e\super{-q\times\tau}, with  q
ranging from 0 --  10. A q value of zero is  exactly what our original
models  had been  doing, namely  not attenuating  \lya~by  dust.  This
models the  dust geometry  of clumpy clouds  in an  empty inter--cloud
medium, i.e.  Neufeld (1991) and Hansen \& Oh (2006).  A q value of 10
attenuates \lya~10 times  more than the continuum.  Thus,  q $\geq$ 10
models  the  effects  of  a  uniform  dust  distribution,  attenuating
\lya~much  more  than  the  continuum  due  to the  fact  that  it  is
resonantly  scattered.  Values  in between  0 and  10  model differing
geometries, with the case of q  = 1 modeling a geometry where \lya~and
the continuum  are attenuated  at the same  rate (possibly  a scenario
with dust in clouds but the inter-cloud medium is not entirely empty).
Introducing  this  parameter  vastly  increased  the  quality  of  the
best-fits, but  it did  come at  a cost.  Since  we had  no additional
data, we could not add another free parameter without taking away one.
We decided to  remove metallicity as a free  parameter, as the effects
of  changing metallicity  had  the  least effect  on  the models.   We
decided to fix a value of Z = .02Z\sol for all models.

\subsection{Object 1}
Object 1 is best fit by a  5 Myr stellar population with a mass of 3.0
$\times$ 10\super{8} M\sol, a  continuous star formation rate, and 1.1
magnitudes of dust extinction (A\sub{1200}) which attenuates \lya~50\% more than the
continuum  (q  =  1.5).   The  quality  of this  fit  was  high,  with
$\chi_{r}^{2}$ =  2.00.  This object is  thus very young,  but its red
color  (specifically $z'$  -  3.6$\mu$m) indicates a  significant
amount  of dust.   However, an  object this  young, especially  with a
continuous SFR,  already has  a pretty high \lya~EW.   Thus even
though it  is best-fit by  a significant amount  of dust, it  does not
require this  dust to enhance its EW,  as evidenced by its  value of q
$>$ 1.  This  object fits the mold of a  ``typical'' \lya~galaxy as it
is  young  and  low  mass.   This  being said,  its  dust  content  is
intriguing  as  it  indicates  that  this  object  is  definitely  not
primitive.

\subsection{Object 2}
Object 2 is  also best fit by a young,  low-mass stellar population (3
Myr; 3.2  $\times$ 10\super{8} M\sol), however it  has a fast-decaying
SFR, with  $\tau$ = 10\super{6}  yr.  It has  less dust than  object 1
(A\sub{1200}  = 0.9  mag),  but  a larger  value  of q,  with  q =  2,
indicating that \lya~is  being extinguished at twice the  value of the
continuum.  However, we can be somewhat skeptical about these results,
as  this   object  has  the  worst   fit  out  of   our  sample,  with
$\chi_{r}^{2}$  =  5.269.   That  being  said,  these  results  aren't
astonishing,  as  they indicate  that  this  object  is possibly  very
similar to object 1.

\subsection{Object 4}
Object 4 is the most intriguing object in our sample.  It has the best
fit of the whole sample, with $\chi_{r}^{2}$ = 1.326, lending the most
credibility to its results.  It is best fit by an 800 Myr, 6.5 $\times$
10\super{9} M\sol stellar population with a continuous SFR.  While the
amount of dust extinction is lower than the other objects (A\sub{1200}
= 0.4 mag)  it has a value of q  = 0, the only object  with q $<$ 1.5.
While 0.4 mag of dust isn't a  lot, if it isn't attenuating \lya, as q
= 0  attests to, it will increase  the EW by $\sim$  50\%, making this
object a candidate  for dust enhancement of the  \lya~EW.  This object
definitely does not fit the mold of a typical \lya~galaxy with its old
age and  larger mass.  Although we  only have one object  like this in
our small sample, its existence means that \lya~galaxes may not all be
uniformly young.

\subsection{Object 6}
Similar to  objects 1 \& 2,  object 6 is  also best-fit by a  young (3
Myr) and  low mass (9.9 $\times$ 10\super{8})  stellar population.  It
has an exponentially decaying SFR with $\tau$ = 10\super{6} yr and 1.8
mag  of  dust attenuating  \lya~50\%  more  than  the continuum.   The
best-fit model ($\chi_{r}^{2}$ = 2.121) has a low EW of $\sim$ 61 \AA,
which is very  comparable to the measured EW of  the object, which was
$\sim$  56   \AA.   Although  this   object  has  a  young   age,  its
fast-decaying SFR and large amount  of dust which {\it is} attenuating
\lya~are keeping the \lya~EW down.  Figure 5 shows the best-fit models
for each object, and Table 4 details the best-fit parameters.

\subsection{Best-Fit Models Without Dust}
As an  exercise, we wanted  to see what  the best-fit models  to these
objects would  be if  we forced the  amount of  dust to be  zero while
keeping  everything else  the  same.   In doing  this,  the number  of
degrees of freedom became three,  because both dust and q were dropped
as  free   parameters,  thus   the  reduced  $\chi^{2}$   reported  is
$\chi^{2}$/3.  Figure 6  shows the results of this  analysis.  With no
dust, all of  the objects were forced to have an  older age to explain
the red colors.  In  all cases, the best fit model with  no dust had a
significantly higher  $\chi^{2}_{r}$ than the  best-fit model allowing
dust,  thus we can  be reasonably  sure that  these objects  have some
dust, and that we have treated it correctly in the models.

\subsection{Monte Carlo Analysis}

To  determine confidence  regions  for our  model  parameters, we  ran
10\super{4} Monte  Carlo simulations.  In  each simulation, we  add or
subtract flux to each band for  each object equal to a gaussian random
deviate multiplied by  the object's flux error in  a given band.  Each
band had its own random  number in each simulation, which was computed
by    the    IDL    function    $\it   RANDOMN$,    which    generates
Gaussian--distributed random numbers.   For each possible model (25200
possible models; 5 SFRs, 24 ages, 21 dust optical depths and 10 values
of q),  the $\chi^{2}$  was computed for  each simulated  object flux.
The result was that for each object, we now had 10000 best-fit models.
In  an  ideal case,  these  models  would  be distributed  around  the
best-fit parameters to the actual observations.

In order  to see  if this  was true, we  plotted contours  showing the
density of best-fit  parameters for each object in  three planes: Dust
vs.   Age, Dust  vs.   q  and Age  vs.   q (Figure  7)\footnote[3]{We
  acknowledge that  we are prevented  from observing part of  the dust
  vs.  q plane due to  selection effects.  A galaxy with A\sub{1200} =
  2.0 and q = 10 would have its \lya~much too supressed to be selected
  via the narrow-band.  Similarly, a galaxy with A\sub{1200} = 2.0 and
  q =  0 would be  bright in the  narrow-band, but its  continuum flux
  would be  vastly attenuated, thus an  object such as  this would not
  have enough  data to be used  in a stellar  population study.}.  The
blue,  green and  red contours  show the  1$\sigma$  (68\%), 2$\sigma$
(95\%)  and 3$\sigma$  (99.5\%) confidence  levels  respectively.  For
object 1,  it appears  as if the  age is  well constrained to  be very
young as the 1$\sigma$ contour is fairly thin along the age axis.  The
same is true for q, constraining it  to be between 1 and 3, ruling out
dust enhancement of  the \lya~EW for this object.   The only parameter
that may be degenerate is dust,  as the 1$\sigma$ contour is very long
along the dust axis.  However, in  no cases is any model best-fit with
zero  dust, and  the  1$\sigma$  contour is  roughly  centered on  the
best-fit value of A\sub{1200} = 1.1.

Object  2  is very  similar  to  object 1  in  that  its  age is  well
constrained to  be young, and it's  q value is well  constrained to be
greater  than  one.  Similarly,  the  dust  in  object 2  is  slightly
degenerate, but  it is  well-constrained to have  some dust,  with the
best-fit value being the most likely  value.  The age of object six is
also  well-constrained  to  be  young,  although  there  is  a  slight
1$\sigma$ blip at t = 100 Myr.  The q value is even better constrained
than in objects  1 or 2, being very centered on  the best-fit value of
1.5.  Dust is still degenerate, although  less so than in objects 1 or
2, with 1$\sigma$ values ranging from 1.5 -- 2.0 magnitudes.

Object 4 is the only object in our sample which shows evidence of dust
enhancement of the  \lya~EW.  The age is mostly  constrained to be $>$
100  Myr, with  a large  1$\sigma$ area  centered around  the best-fit
model at 800 Myr, and a  smaller 1$\sigma$ area around 200 Myr (and an
even smaller  one at 30 Myr).   A range of dust  values are permitted,
but generally  lie in  the range  from 0.1 --  0.9 magnitudes.   The q
parameter is  definitely constrained to  be $<$ 1, with  the 1$\sigma$
area centered  around the best-fit, with  no part of  it being greater
than q  = 0.25.  Thus this  object is well-constrained to  have an old
age and a small amount of dust with a dust geometry which is enhancing
the  \lya~EW.  It  should be  noted  that the  effect of  dust is  not
enormous.  Dust extinction of 0.4 magnitudes will increase the \lya~EW
by $\sim$ 50\%, meaning that the  EW of this object was not negligible
beforehand, namely  due to a  constant SFR.  However, the  dust pushes
this object's best-fit  model EW from a value of  $\sim$ 120 \AA~up to
$\sim$ 180  \AA.  Given  the model  EW curves we  saw earlier,  if one
ignored the possible effects of dust one would expect to see many more
120 \AA~EW objectss  than 180 \AA.  While this effect  is only seen in
one object, it  shows that we cannot afford to  ignore the effect dust
could  have on  \lya~EWs, as  we have  shown it  can help  explain the
perceived over-abundance of high-EW sources.

\section{Discussion}

\subsection{\lya~Detection Fraction}
Out of  the whole GOODS  CDF-S catalog, we  found that there  were 229
B-band  dropouts.   The  dropout  criteria  selects  galaxies  with  a
$\Delta$z range of $\sim$ 1.0.  The width of our narrow-band filter is
80 \AA~around  a central  wavelength of 6563  \AA, corresponding  to a
$\Delta$z range of 0.07 for \lya.   Thus, about 0.07 $\times$ 229 = 16
B-dropouts  lie in  the redshift  range  where we  could detect  \lya.
Among these, only five show a narrow-band excess.  This corresponds to
a \lya~detection  fraction of  31\%, comparable with  previous studies
(e.g. Shapley et al. 2003; 25\%).

\subsection{\ha~Equivalent Widths}
When comparing the  best-fit models to the data  points, it is obvious
that the model \ha~flux is  dominating the model 3.6$\mu$m flux, as is
evident by  the fact that  the 3.6$\mu$m data  point is far  above the
model continuum  level.  The model rest-frame \ha~EWs  are 3800, 8500,
200 and  1700 \AA~for objects  1, 2, 4  and 6 respectively.   At first
glance, these seem large, but one  needs to remember that with the
exception of object  4, these objects have been  found to be extremely
young, low-mass,  star-forming galaxies.  Keeping that  in mind, these
values of \ha~EW are quite reasonable.   

In a survey of HII regions  in a nearby galaxy, Cedres, Cepa \& Tomita
(2005) found that  the \ha~EW distribution peaked at  $\sim$ 1000 \AA,
with a few HII regions with EWs  as high as 10000 \AA.  In an analysis
of a z=6.56 \lya~galaxy, Chary et al. (2005) derived an \ha~EW of 2000
\AA,  concluding that  the \ha~line  dominated the  4.5$\mu$m  flux in
their object.   While computing the synthetic  properties of starburst
galaxies,  Leitherer and Heckman  (1995) computed  a \ha~EW  of $\sim$
3200 \AA~for a population with  0.1Z\sol and a constant star formation
rate.  Taking  into account model differences  (including a difference
of 5X in metallicity) we find our values of the \ha~EW consistent with
those previosuly found in the literature for star forming galaxies.

\subsection{Comparison to Other Work}

It is  useful to compare  our results to  those of other  studies done
using {\it HST} and {\it Spitzer} data together, as the IR data (which
corresponds to  the rest-frame optical  at z = 4.4)  better constrains
the  stellar  masses  as  it  is  less attenuated  by  dust  than  the
rest-frame UV.   Two recent studies have been  published studying LAEs
at z $\sim$ 5 and z $\sim$ 5.7 by Pirzkal et al. (2007) and Lai et al.
(2007) respectively.  Pirzkal et  al.  studies the stellar populations
of nine LAEs in the HUDF  detected on the basis of their \lya~emission
lines  in the  Grism ACS  Program for  Extragalactic  Science (GRAPES;
Pirzkal et  al.  2004;  Xu et al.  2007; Rhoads  et al. 2007  in prep.)
survey.  Using  similar stellar population models as  ours, they found
very young  ages ($\sim$  10\super{6} yr) and  low masses  ($\sim$ few
$\times$  10\super{7} M\sol),  with some  dust in  a good  fraction of
their  objects.  Lai et  al. studied  three narrow-band  selected LAEs
spectroscopically confirmed to lie at  z $\sim$ 5.7 in the GOODS HDF-N
field.  Also using BC03 models, with an instantaneous burst they found
ages from 5 -- 100 Myr (up to 700 Myr with a constant SFR), and masses
from  10\super{9}  -  10\super{10}  M\sol.  Their  objects  were  also
best-fit by dust, with E(B-V) $\sim$ 0.3 -- 0.4.

Three of our four objects have ages consistent with those from Pirzkal
et al. and  Lai et al., who find  ages from $\sim$ 1 --  100 Myr.  Our
fourth object has an age much larger than those found in these papers,
at 800 Myr.  Eight out of the  nine Pirzkal et al.  objects have M $<$
10\super{8} M\sol,  whereas the  Lai et al.   objects have  masses two
orders of magnitude  higher.  The masses of our  objects reside in the
middle, from 10\super{8} -  10\super{9} M\sol.  The likely reason that
our masses  are higher than those  of Pirzkal et al.   is because they
reach fainter  luminosities, detecting objects with  emission lines as
faint as  5 $\times$  10\super{-18} erg cm\super{-2}  s\super{-1}.  On
the flipside,  the larger masses  from Lai et  al.  are likely  due to
their selection  criteria.  They found  12 LAEs in the  GOODS-N field,
but they only analyzed the  three objects with significant 3.6 and 4.5
$\mu$m flux.   They acknowledge that  their masses are high,  and that
this may  be due to  selecting objects from  the high-mass end  of the
mass distribution.

Another  difference is  how  each study  treated  \lya~in their  model
fitting.  Pirzkal et  al. chose to exclude the ACS  i' band from their
fitting, as at their redshifts  the \lya~line would fall in this band,
and they didn't want to include the uncertainties that subtracting the
line flux from  the i' flux would create.  Lai et  al.  chose to treat
\lya~by by  estimating the amount  of flux \lya~contributed to  the i'
flux, then adding  this percentage as an i'-band  error ($\sim$ 30\%).
As described  in detail in earlier  sections, we chose  to add the
\lya~emission to  the models in  order to investigate the  effect dust
had on the \lya~EW.  While each of these methods has its merit, the
method  of Pirzkal  et  al.   likely introduces  the  least amount  of
uncertainty by simply not  including the broad-band which contains the
\lya~emission.  They  had the luxury that  in addition to  the ACS and
IRAC data, they  also had {\it HST}/NICMOS and  VLT/ISAAC NIR data, so
cutting a  band from  the fitting did  not force  them to throw  out a
constraint.  

One other  intersting study is  that done by  Chary et al.   (2005) by
fitting stellar population  models to a z =  6.56 \lya~galaxy with NIR
and  IRAC data.   As we  metioned  above, they  determined that  their
4.5$\mu$m  flux was dominated  by a  \ha~line, implying  vigorous star
formation.   Taking this  line  into account,  they  found a  best-fit
stellar  population of  5  Myr, 8.4  $\times$  10\super{8} M\sol  with
A\sub{V} =  1.0 mag.  This is  even dustier than  our objects, showing
that not only is dust common at z = 4.4, it may exist in \lya~galaxies
up to z $\sim$ 6.5.

By including \lya~in our models rather then attempting to subtract out
its influence in the observed broad-bands, we were able to investigate
the  effect dust geometry  had on  the \lya~EW.   While the  other two
studies  listed  here did  not  do  this  analysis, they  did  include
homogeneous dust in  their models.  They found that  many/all of their
objects were  best fit by including  at least a small  amount of dust.
This confirms  our result  that regardless of  the effect of  the dust
geometry, one needs to include dust when analyzing LAEs, as it appears
to be present  in a majority of them.  Failure to  include dust in the
models  will result  in a  poorer fit,  likely with  an older  or more
massive stellar population than is really in place.

\section{Conclusion}

We  have presented  the results  of our  analysis of  four narrow-band
selected \lya~galaxies  in the  GOODS CDF-S.  By  observing in  such a
well-studied field, we were able  to take advantage of the plethora of
deep public  broadband data  (GOODS {\it HST}/ACS,  {\it Spitzer}/IRAC
and  MUSIC),  as  well   as  previous  spectroscopic  studies  (PEARS)
available  in  the  region.   Three  out  of  our  four  objects  were
spectroscopically confirmed to lie at z = 4.4.  The other object has a
photometric redshift of 4.42.

Previously,  it  has  been  assumed  that  \lya~galaxies  were  young,
low-mass galaxies  with little  to no dust.   Even when dust  has been
found, it  has been assumed  to be homogeneously  distributed, meaning
that it  would attenuate  the \lya~flux much  more than  the continuum
flux  because  \lya~photons  are  resonantly  scattered.   However,  a
scenario has been theoretically studied  where the geometry of a dusty
ISM could  actually enhance the  \lya~EW rather than reduce  it.  This
scenario consists of an ISM where the dust resides in clouds, mixed in
with  neutral hydrogen,  whereas  the inter-cloud  medium  is hot  and
ionized.  In this scenario,  \lya~photons would effectively bounce off
of the  surface of the  clouds, finding it  much easier to  escape the
galaxy than  continuum photons.   Thus a galaxy  could exhibit  a much
larger \lya~EW than intrinsic to the underlying stellar population.

In  order to  investigate this,  we have  compared the  fluxes  of our
objects to stellar population  models.  We have included \lya~emission
in the models,  as well as introducing a  geometry parameter, q, which
dictates how much \lya~is attenuated by a given dust amount.  Three of
our four objects were best-fit by models with a very young age ($\sim$
few  Myr) and  low  stellar masses  ($\sim$  few $\times$  10\super{8}
M\sol), with a  significant amount of dust (A\sub{1200}  = 0.9 -- 1.8)
attenuating the  \lya~flux more than the  continuum (q =  1.5 -- 2.0).
However,  one object,  object 4,  is  best-fit by  a vastly  different
model.  This object is best fit  by a much older, more massive stellar
population, with  an age of 800  Myr and M =  6.5 $\times$ 10\super{9}
M\sol.  What is interesting about this  object is that its EW is among
the  highest in  our sample.   Even with  a constant  SFR, an  800 Myr
stellar population  would not be able  to produce a  \lya~EW above 120
\AA, while this object has an observed rest-frame EW of $\sim$ 150 \AA
and a best-fit model EW of  $\sim$ 180 \AA.  Thus something is causing
the EW of this  object to be enhanced, and the models  tell us that it
is  due  to  dust.  This  object  is  best-fit  by A\sub{1200}  =  0.4
magnitudes  of dust,  and while  this is  less than  any of  the other
objects, the geometry parameter is best fit to be q = 0.0.  This means
that the 0.4 magnitudes of dust are attenuating the continuum, but not
the \lya~flux, enhancing the EW by  50\%. Figure 7 shows a fair margin
of  parameter space  is allowed  for most  objects at  the 2-3$\sigma$
level.

Comparing our  results to those from  other studies, we  find that our
age  and   mass  results  are  consistent  when   data  and  selection
differences  are  accounted  for.   However,  no  other  observational
studies have yet tried to model the effect dust could have on \lya~for
differing geometries.  While  our sample size is small,  analysis of a
larger  sample could  indicate that  there are  really two  classes of
\lya~galaxies:  a young,  low mass  \lya~galaxy with  an intrinsically
high \lya~EW and  an older, higher mass galaxy  with a lower intrinsic
EW which has been enhanced due to dust.

\begin{acknowledgements}
  Support for  this work  was provided in  part by NASA  through grant
  numbers HST-AR-11249,  HST-GO-10240 and HST-GO-10530  from the SPACE
  TELESCOPE SCIENCE INSTITUTE, which is operated by the Association of
  Universities for  Research in  Astronomy, Inc., under  NASA contract
  NAS5-26555.   This work  was  also supported  by  the Arizona  State
  University (ASU) Department of Physics,  the ASU School of Earth and
  Space Exploration  and the ASU/NASA  Space Grant.  We  thank Russell
  Ryan,  Seth Cohen  and  Evan Scannapieco  for helpful  conversations
  which greatly improved this paper.
\end{acknowledgements}

\input {tab1.tex}
\input {tab2.tex}
\input {tab3.tex}
\input {tab4.tex}

\begin{figure}
\epsscale{1.0}
\plotone{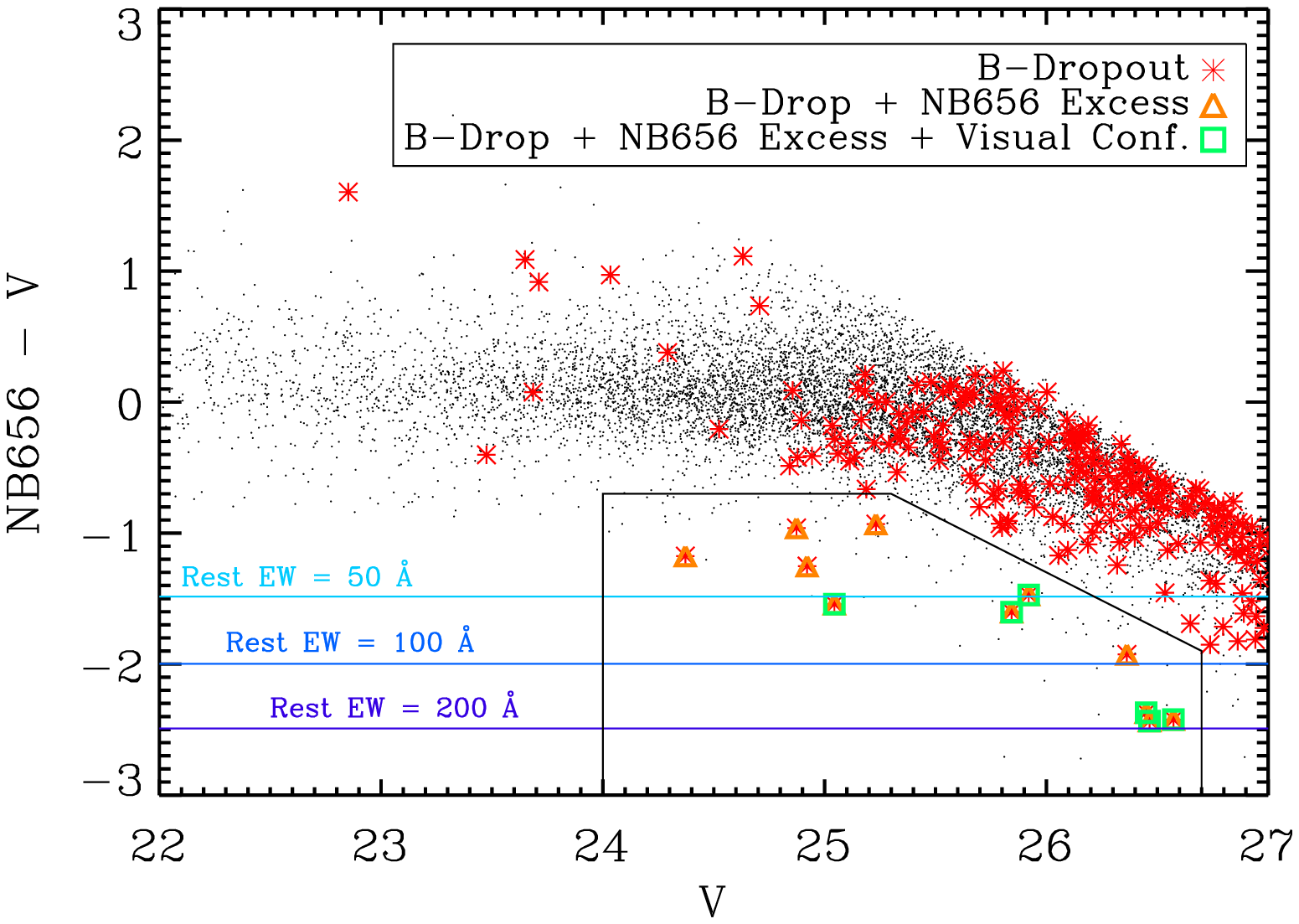}
\caption{Color-magnitude  diagram of  all objects  extracted  from the
  narrow-band (NB656)  image via PSF-fitting.  The 11  objects showing a
  narrow-band excess  that were also  B-dropouts were selected as  z =
  4.4 \lya~galaxy candidates.  Six of these 11 initial candidates were
  confirmed following a visual inspection of the image, with the other
  five being either too close to a neighboring bright star or visually
  undetected  in the  image.  Also  shown in  this plot  are  lines of
  constant rest-frame equivalent width for three values of EW.}
\end{figure}
\begin{figure}
\epsscale{1.7}
\plottwo{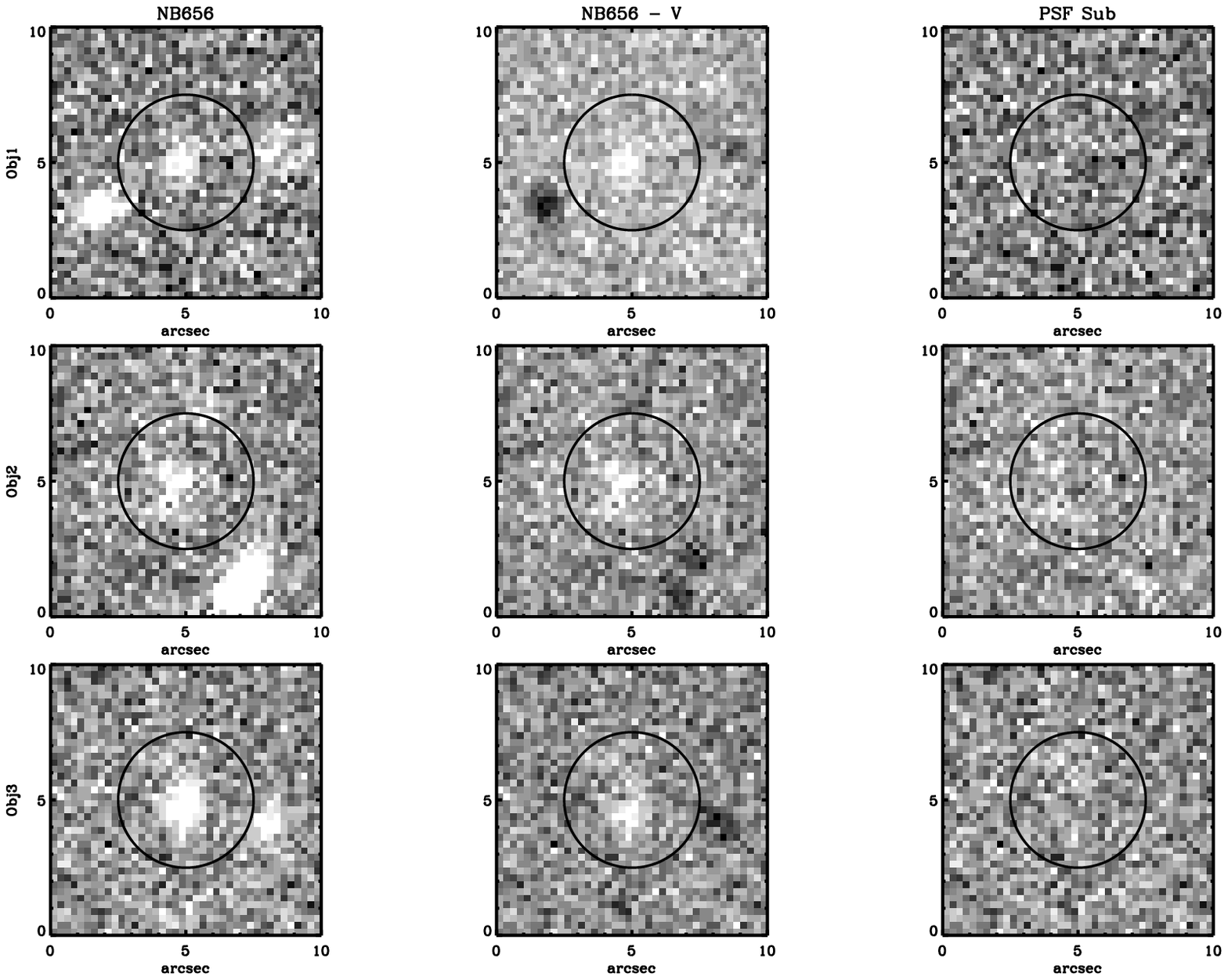}{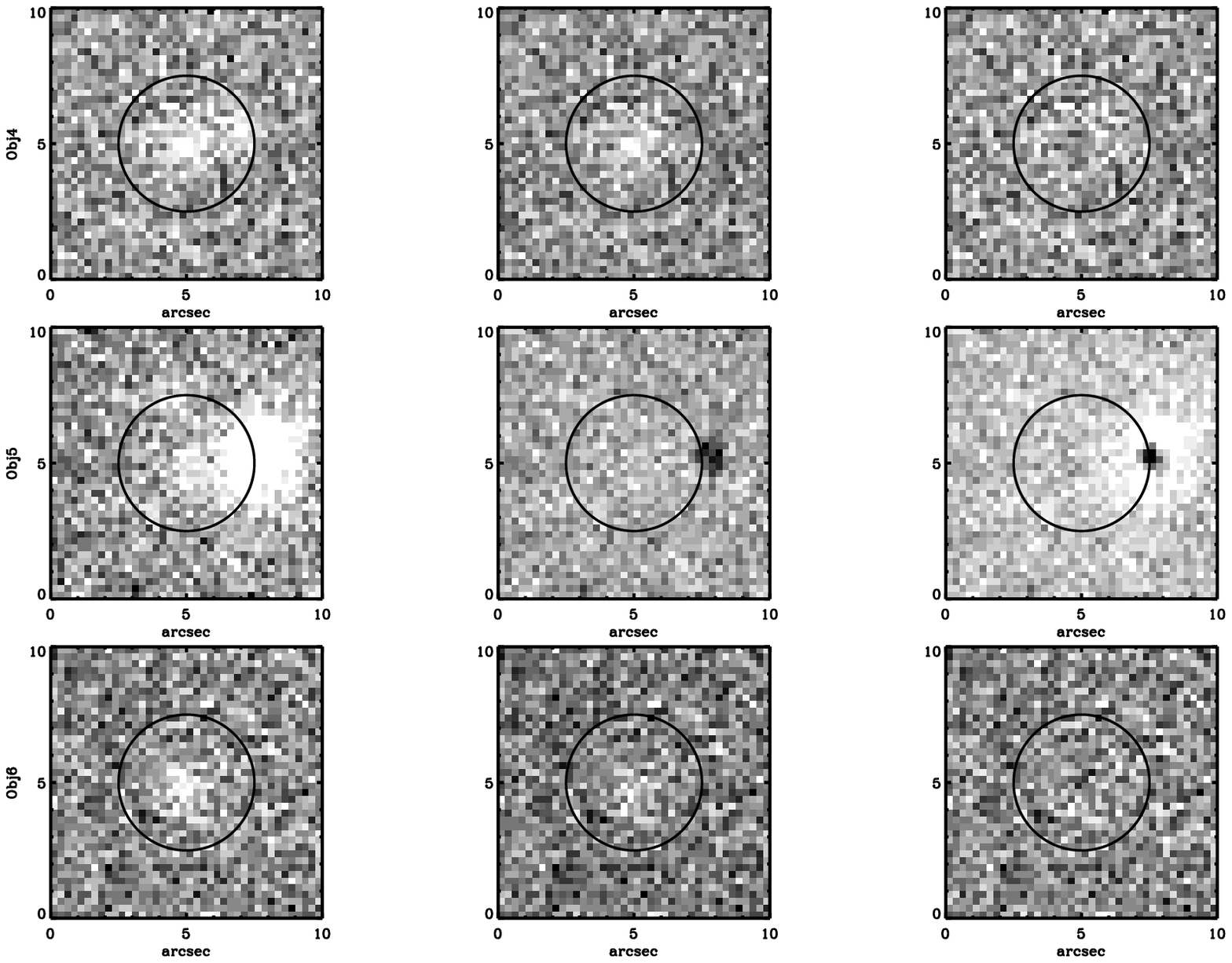}
\caption{Results  from  our  tests   to  check  the  validity  of  our
  candidates.   The  first  column   shows  a  10\arcs~stamp  of  each
  candidate  in  the  NB656  image.   The  second  column  shows  each
  candidate in a narrow-band excess (NB656 - V) image.  We see flux in
  five out of the six objects, visually confirming that they exhibit a
  narrow-band excess.  The one object that doesn't definitively show a
  narrow-band excess  is very near  a bright star, which  could affect
  this  observation.  Regardless,  this object  is thrown  out  of our
  analysis as  the nearby star contaminates its  {\it Spitzer} fluxes.
  The third column shows  each candidate's position in the subtraction
  (PSF-residual) image.  The  lack of any flux above  the noise proves
  that the PSF extraction was satisfactory.  }
\end{figure}
\begin{figure}
\epsscale{1.0}
\includegraphics[scale=1.20,angle=90]{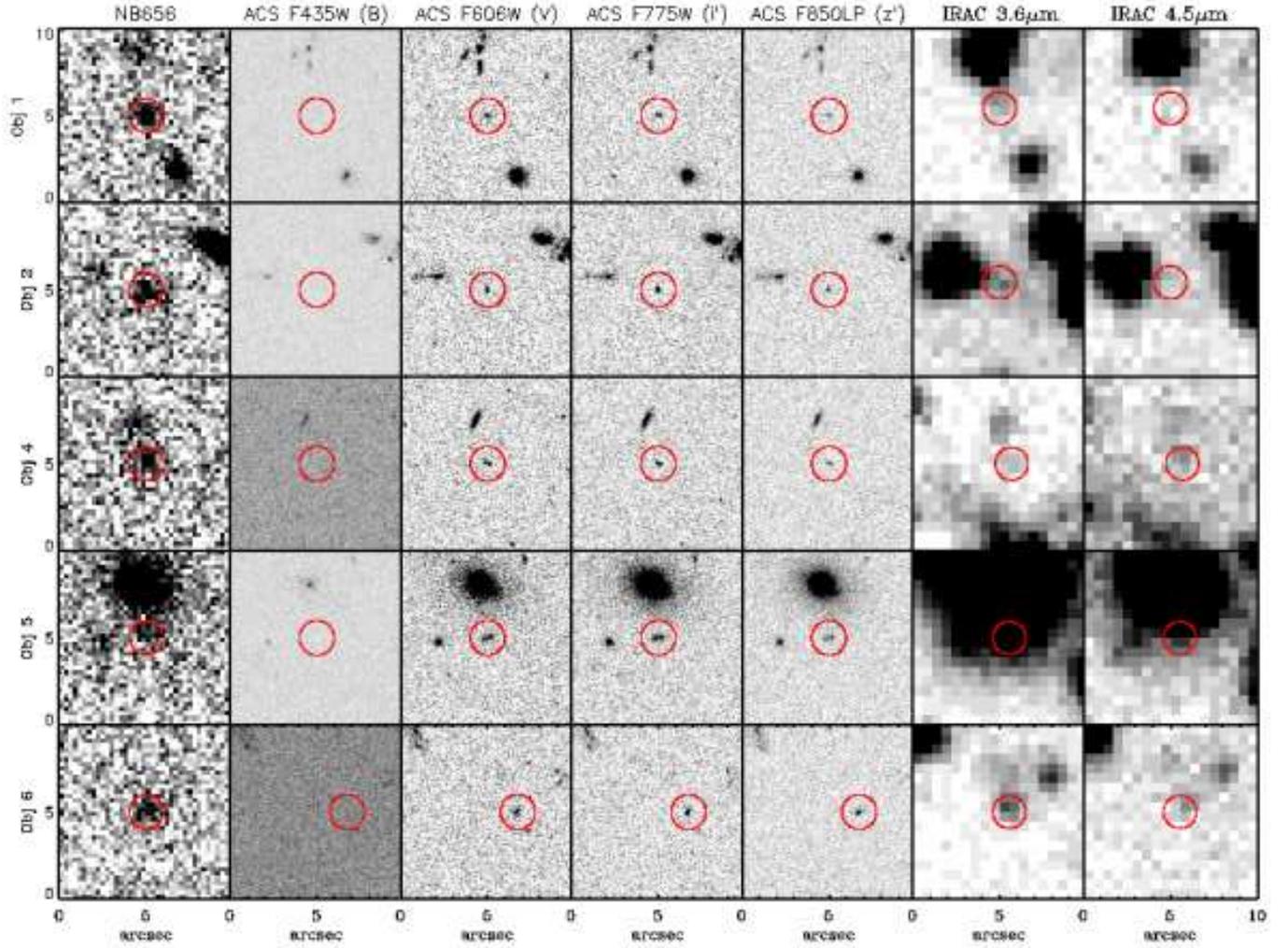}
\caption{10\arcs~stamps of each candidate in the narrow-band, the four
  {\it HST}/ACS bands, and  two IRAC/{\it Spitzer} bands.  The circles
  are 2\arcs~in diameter, and are drawn to highlight the objects.}
\end{figure}
\begin{figure}
\epsscale{1.0}
\plotone{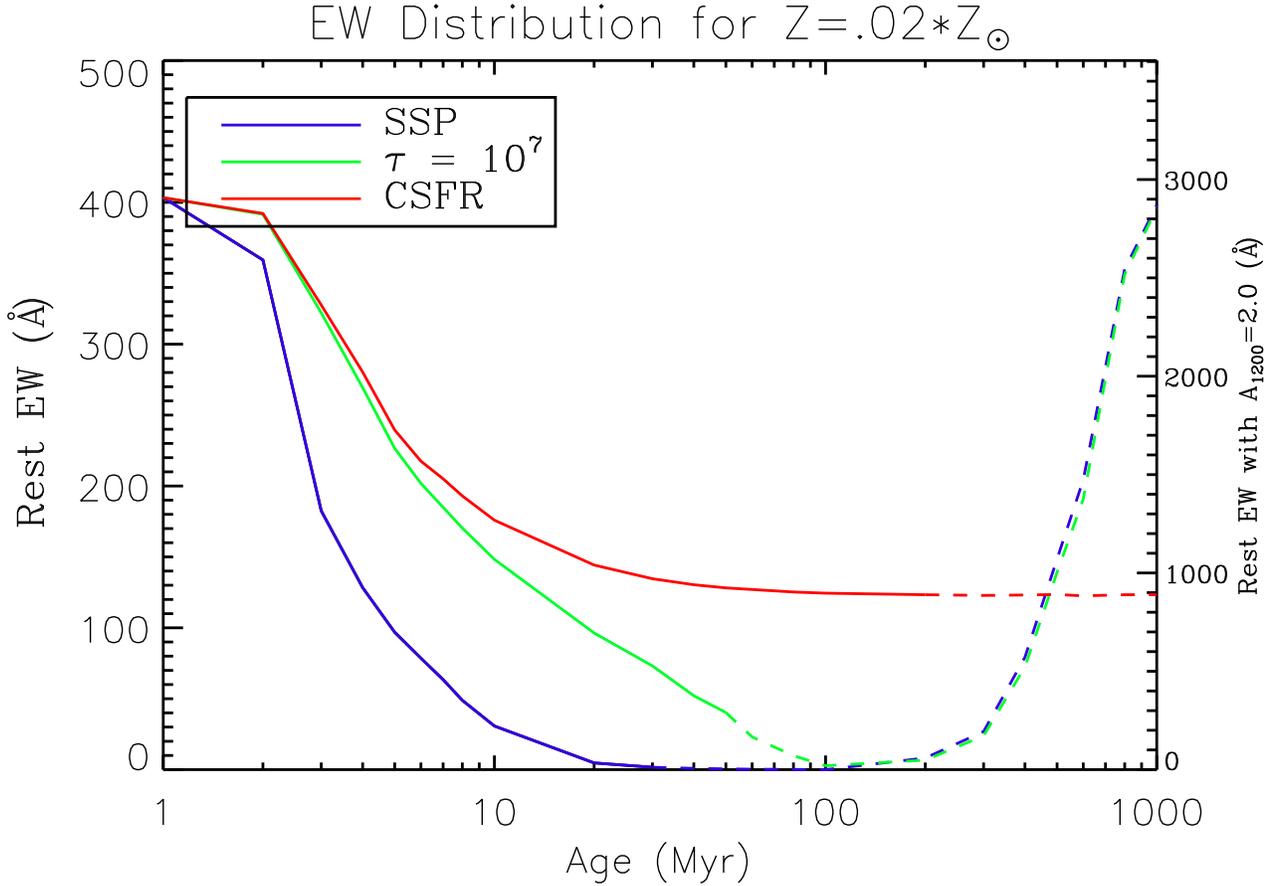}
\caption{Equivalent  width  distribution  from  the  models  for  Z  =
  .02Z\sol.  The  upturn in  EW at  late age is  caused by  either hot
  horizontal branch  stars and/or planetary nebulae  nuclei, either of
  which will  dominate the  UV flux at  later ages.  However,  at this
  point the total UV flux is much less that it was at early ages, such
  that a stellar population this old  would likely not be picked up in
  a narrow-band selected survey.   Due to this, the line-style changes
  to dashed at the point where the V-band flux drops below 10\% of its
  value at 10\super{6} years.  The left-hand axis shows the rest-frame
  EW with  no dust,  and the right-hand  axis shows the  rest-frame EW
  with A\sub{1200} = 2.0.  In order to observe a galaxy with an EW $>$
  200 \AA, it either  needs to be only a few Myr  old with no dust, or
  with  dust it  can be  older than  10 Myr.   Comparison  to previous
  studies (Charlot \&  Fall 1991; Malhotra and Rhoads  2002) shows that
  our model EWs are consistent with previous results.}
\end{figure}
\begin{figure}
\epsscale{1.15}
\plotone{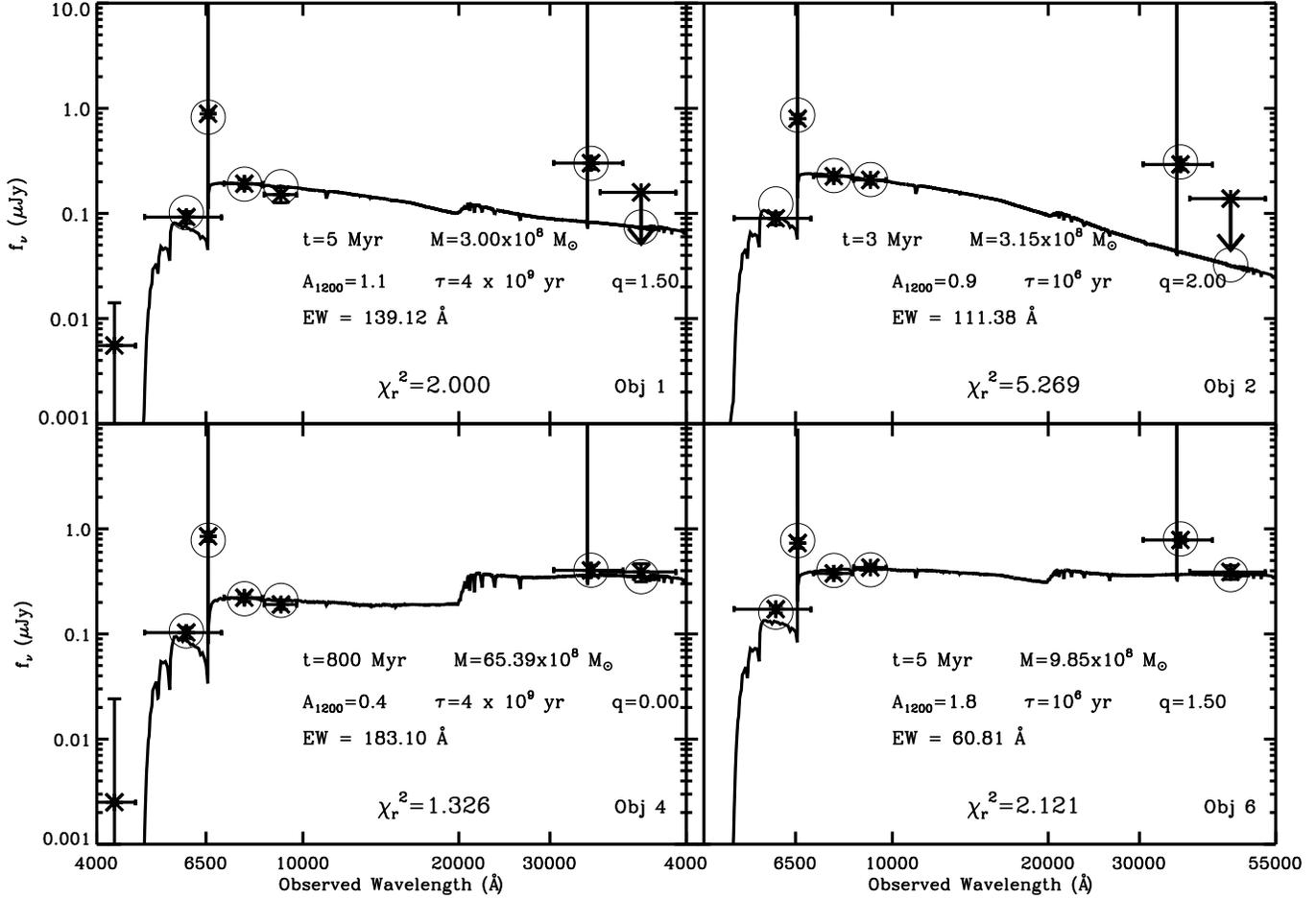}
\caption{The  best-fit model  spectra, with  the observed  data points
  overplotted.  The open circles are  centered on the positions of the
  bandpass averaged fluxes  of the best-fit models.  Object  1 is best
  fit with a young, low-mass stellar population with a continuous star
  formation  rate.  The best-fit  model contains  over a  magnitude of
  dust, but  its geometry is  such that it is  extinguishing \lya~50\%
  more than the continuum.  Object  2 is best-fit by a young, low-mass
  stellar population with an  exponentially decaying SFR with $\tau$ =
  10\super{6} yr.   There is a little  less than a  magnitude of dust,
  but it is  extinguishing \lya~at twice the amount  of the continuum.
  Object 4 is by far the most interesting object in our sample.  It is
  best  fit by  a much  older, higher-mass  stellar population  with a
  continuous SFR.   It has less dust  that the other  models, but this
  dust is not attenuating \lya~at all, enhancing the observed \lya~EW.
  The combination of dust and a continuous SFR is allowing this object
  to have a high EW  at an age of 800 Myr.  Object  6 is best-fit by a
  young,  low-mass model  similar to  objects  1 \&  2.  However,  the
  best-fit  model  has  nearly  two  magnitudes  of  dust  attenuating
  \lya~twice as much as the continuum, resulting in a lower EW.  }
\end{figure}
\begin{figure}
\epsscale{1.00}
\plotone{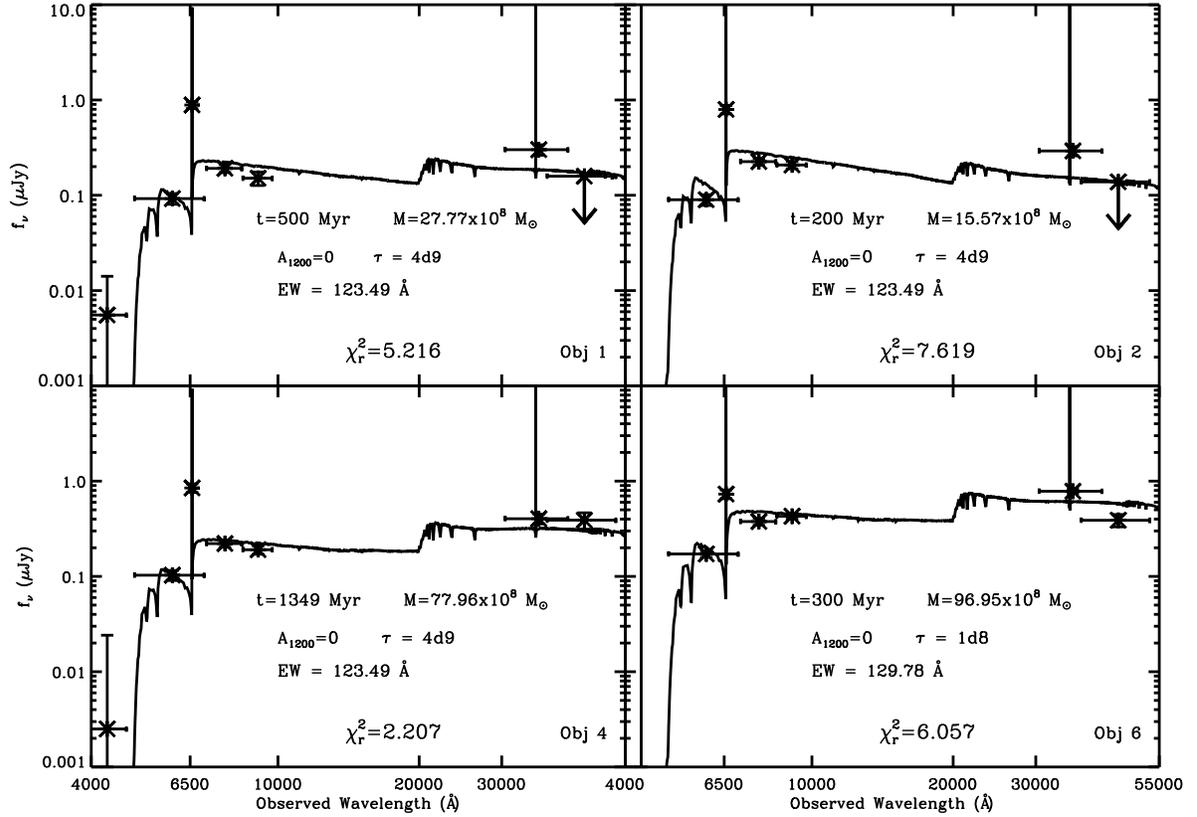}
\caption{The best-fit  stellar populations forcing the  models to have
  no dust.   All objects are  best-fit by older, more  massive stellar
  populations, but the quality of these fits are of much lower quality
  (higher $\chi^{2}$) than the best-fit models with dust.}
\end{figure}
\begin{figure}
\epsscale{1.6}
\plottwo{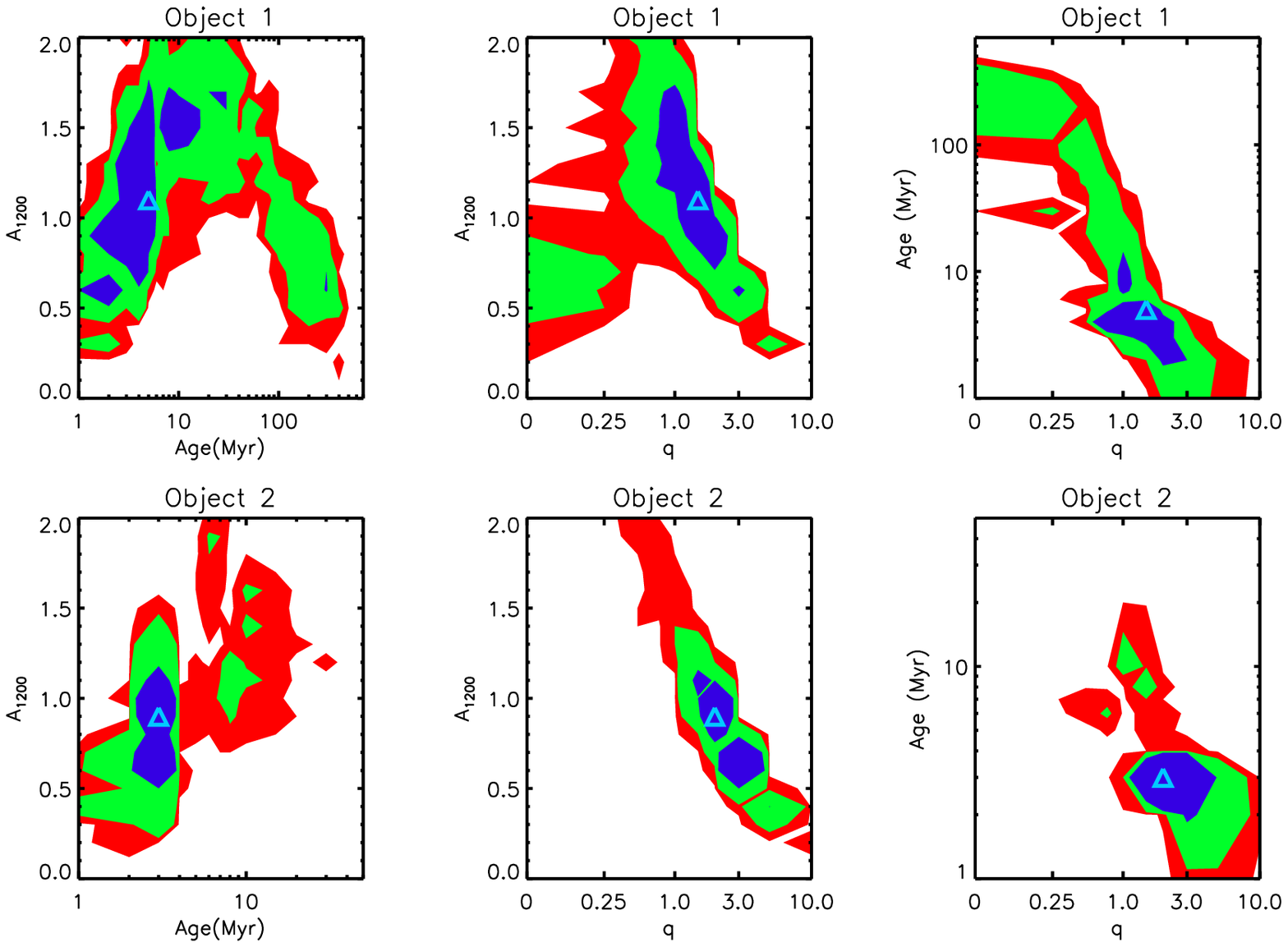}{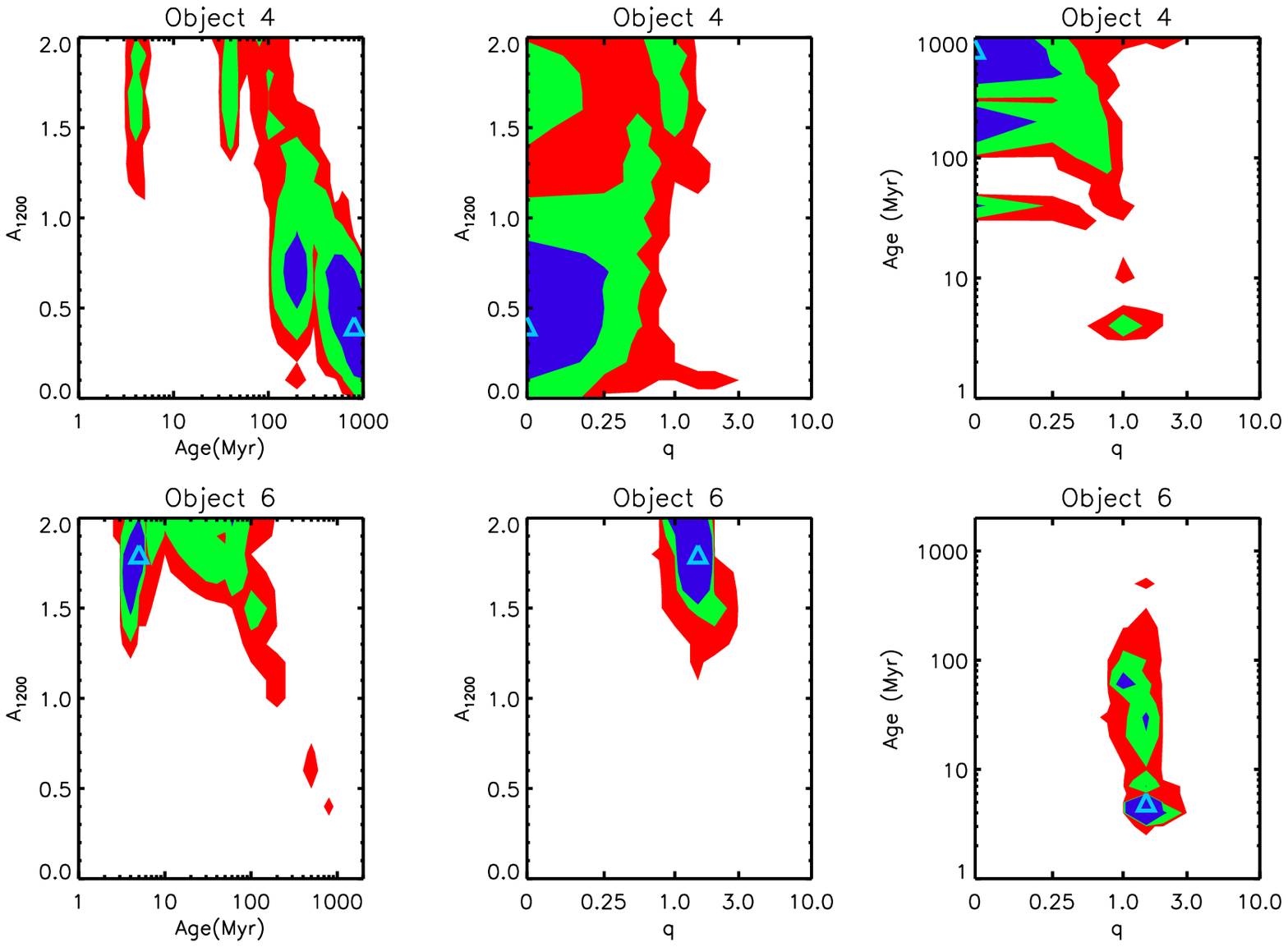}
\caption{Contours showing  the density of  results of our  10000 Monte
  Carlo  simulations.  The best-fit  model from  the actual  object is
  denoted  by the blue  triangle.  The  four objects  are in  the four
  rows,  and  the  columns  represent  three  different  planes:  Dust
  vs. Age, Dust  vs. q and Age vs.  q (note that the range  of the age
  axis is different  for some of the objects).   67\% of the best-fits
  from the simulations  fall in the blue contour, and  as such this is
  the 1$\sigma$ contour.  Likewise, the green is the 2$\sigma$ contour
  and red is  the 3$\sigma$ contour.  Studying these  figures, we were
  satisfied  that  our  best-fit  models  accurately  represented  the
  most-likely best-fit for each object.}
\end{figure}

\end{document}

%% file: tab1.tex

\begin{deluxetable}{ccccc}
\tabletypesize{\small}
\tablecaption{Ly$\alpha$ Galaxy Candidates}
\tablewidth{0pt}
\tablehead{
\colhead{Name} & \colhead{m$_{AB}$ (NB656)} & \colhead{m$_{AB}$ (F606W)} & \colhead{Rest EW (\AA)} & \colhead{Spectroscopic Confirmation}\\
}
\startdata
Object 1&24.03&26.49&189.7 $\pm$ 46.8&$\sim$4.4\\
Object 2&24.15&26.52&166.6 $\pm$ 42.9&$\sim$4.4\\
Object 3&23.50&24.84&52.8 $\pm$ 5.9$^{*}$&AGN @ z=3.19\\
Object 4&24.08&26.37&149.1 $\pm$ 36.4&No Spectra\\
Object 5&24.48&25.89&46.9 $\pm$ 13.0&No Spectra\\
Object 6&24.24&25.81&56.3 $\pm$ 11.0&$\sim$4.4\\

\enddata
\tablecomments{$^{*}$ Object 3 is a known AGN at z $=$ 3.19, and the line we detected in the narrow-band filter is [CIV], thus the EW for object 3 is the [CIV] rest-frame EW for an object at z = 3.19.  The calculated rest-frame EWs for the other objects assume a redshift equal to that of \lya~at the center of the \ha~filter, which is 4.399.  Objects 1, 2, 3 and 6 had grism spectra from the PEARS survey, thus we were able to spectroscopically confirm objects 1, 2 and 6 to be at z $\approx$ 4.4.}

\end{deluxetable}


%% file: tab2.tex

\begin{deluxetable}{ccccccccc}
\tabletypesize{\small}
\rotate
\tablecaption{Ly$\alpha$ Galaxy Candidates}
\tablewidth{0pt}
\tablehead{
\colhead{Name} & \colhead{IAU Format Name} & \colhead{m$_{NB656}$} & \colhead{m$_{B}$} & \colhead{m$_{V}$} & \colhead{m$_{i'}$} & \colhead{m$_{z'}$} & \colhead{m$_{Ch1}$} & \colhead{m$_{Ch2}$}\\
}
\startdata
Obj 1&J033215.991-274231.57&24.03$\pm$0.09&29.54$\pm$ 1.33&26.49$\pm$0.14&25.70$\pm$0.10&25.95$\pm$0.17&25.20$\pm$0.14&25.64$\pm$0.36$^{*}$\\
Obj 2&J033239.771-275114.95&24.15$\pm$0.11&98.95$\pm$98.91&26.52$\pm$0.14&25.52$\pm$0.09&25.61$\pm$0.10&25.24$\pm$0.12&25.92$\pm$0.36$^{*}$\\
Obj 4&J033258.380-275339.58&24.08$\pm$0.11&30.40$\pm$ 3.10&26.37$\pm$0.13&25.54$\pm$0.09&25.70$\pm$0.11&24.89$\pm$0.13&24.92$\pm$0.21\\
Obj 5&J033224.227-274129.48&24.45$\pm$0.17&98.95$\pm$98.91&25.89$\pm$0.10&25.10$\pm$0.09&24.84$\pm$0.08&23.09$\pm$0.02&24.23$\pm$0.10\\
Obj 6&J033248.244-275136.90&24.24$\pm$0.12&98.95$\pm$98.91&25.81$\pm$0.09&24.96$\pm$0.10&24.82$\pm$0.07&24.16$\pm$0.05&24.93$\pm$0.16\\
\enddata
\tablecomments{$^{*}$   3$\sigma$  upper   limits.    Coordinates  and
  magnitudes for each of  our candidate \lya~galaxies.  Each magnitude
  has been corrected to represent the total magnitude for the object.}

\end{deluxetable}


%% file: tab3.tex

\begin{deluxetable}{ccccc}
\tabletypesize{\small}
\tablecaption{Maximum Possible Age Allowed for a Stellar Population with EW $>$ 200 \AA}
\tablewidth{0pt}
\tablehead{
\colhead{SFR} & \colhead{No Dust} & \colhead{ } & \colhead{A$_{1200}$ = 2.0} & \colhead{}\\
\colhead{ } & \colhead{Z = .02*Z\sol} & \colhead{Z = Z\sol} & \colhead{Z = .02*Z\sol} & \colhead{Z = Z\sol}\\
}
\startdata
SSP&3 Myr (182 \AA)&1 Myr (193 \AA)&10--20 Myr (222--34 \AA)&5--6 Myr (424--181 \AA)\\
$\tau$ = 10$^{7}$&7 Myr (184 \AA)&3 Myr (186 \AA&50--60 Myr (291--167 \AA)&50--60 Myr (300--193 \AA)\\
Constant SFR&8 Myr (193 \AA)&3 Myr (190 \AA)&Never&Never\\
\enddata
\tablecomments{Values in parenthesis are  the equivalent widths at the
  reported ages.  Since  the grid of model ages  is not continuous, in
  some cases there is a large  amount of time around when the EW drops
  below 200 \AA~in the models.  In  these cases we report the ages and
  EWs on either side of 200 \AA.  Depending on the star formation rate
  and metallicity,  two magnitudes of clumpy dust  extinction can keep
  the \lya~EW above  200 \AA~up to 3--20 times  longer than just young
  stars alone (indefintiely for a  population which suffers no drop in
  star formation activity).  Models with dust have q = 0.}

\end{deluxetable}


%% file: tab4.tex

\begin{deluxetable}{ccccccc}
\tabletypesize{\small}
\tablecaption{Best-Fit Physical Parameters}
\tablewidth{0pt}
\tablehead{
\colhead{Name} & \colhead{$\chi^{2}_{r}$} & \colhead{Age} & \colhead{Mass} & \colhead{A$_{1200}$} & \colhead{SFR} & \colhead{q}\\
\colhead{$ $} & \colhead{$ $} & \colhead{(Myr)} & \colhead{(M\sol)} & \colhead{(mag)} & \colhead{$ $} & \colhead{$ $}\\
}
\startdata
Object 1&2.000&5&3.00 $\times$ 10$^{8}$&1.1&Continuous&1.50\\
Object 2&5.269&3&3.15 $\times$ 10$^{8}$&0.9&$\tau$ = 10$^{6}$&2.00\\
Object 4&1.326&800&65.39 $\times$ 10$^{8}$&0.4&Continuous&0.00\\
Object 6&2.121&5&9.85 $\times$ 10$^{8}$&1.8&$\tau$ = 10$^{6}$&1.50\\
\enddata
\tablecomments{The best-fit parameters for each of our objects.  These
  were found  by minimizing the reduced $\chi^{2}$ between each object  and a
  grid of models.}

\end{deluxetable}
